\documentclass[10pt,twocolumn,letterpaper]{article}

\usepackage[top=1in,bottom=1in,left=0.75in,right=0.75in,columnsep=0.25in]{geometry}

\usepackage[T1]{fontenc}
\usepackage{newtxtext,newtxmath}

\usepackage[authoryear]{natbib}

\usepackage{amsmath}
\usepackage{amssymb}
\usepackage{algorithm}
\usepackage{algorithmic}

\usepackage{array}
\usepackage{graphicx}
\usepackage{booktabs}
\usepackage{multirow}
\usepackage{colortbl}
\usepackage{xcolor}
\usepackage{makecell}
\usepackage{subfig}
\usepackage{wrapfig}

\usepackage[most]{tcolorbox}

\usepackage{listings}
\usepackage{enumitem}

\usepackage{etoolbox}
\usepackage{placeins}

\renewcommand{\footnotesize}{\scriptsize}

\usepackage{caption}
\usepackage[colorlinks]{hyperref}
\hypersetup{citecolor={blue!80!black},linkcolor={blue!80!black},urlcolor={blue!80!black}}
\graphicspath{{figures/}}

\definecolor{oursrow}{RGB}{232,245,233}
\definecolor{grayrow}{RGB}{240,240,240}
\definecolor{promptbg}{RGB}{242,247,253}
\definecolor{promptborder}{RGB}{70,130,180}
\definecolor{gaingreen}{RGB}{34,139,34}

\newtcolorbox{promptbox}{
 colback=promptbg, colframe=promptborder,
 boxrule=0.5pt, arc=1mm,
 left=4pt, right=4pt, top=4pt, bottom=4pt,
 breakable, enhanced}

\newcommand{\best}[1]{\textbf{#1}}
\newcommand{\na}{\text{---}}
\newcommand{\gain}[1]{{\textcolor{gaingreen}{\,#1}}}

\begin{document}
\def\floatpagepagefraction{1}
\def\textpagefraction{.001}

\twocolumn[
  \begin{center}
  {\LARGE\bfseries POSIM: A Multi-Agent Simulation Framework for Social Media\\[4pt]Public Opinion Evolution and Governance\par}
  \vspace{14pt}
  {\large
  Yongmao Zhang\quad
  Kai Qiao\quad
  Zhengyan Wang\quad
  Ningning Liang\quad
  Dekui Ma\quad
  Wenyao Sun\quad
  Jian Chen\quad
  Bin Yan\textsuperscript{*}\par}
  \vspace{8pt}
  {\normalsize\itshape Henan Key Laboratory of Imaging and Intelligent Processing,\\
  Information Engineering University, Zhengzhou 450001, PR China\par}
  \vspace{6pt}
  {
  \textsuperscript{*}Corresponding author.\par}
  \end{center}
  \vspace{10pt}

  \centerline{\rule{0.92\textwidth}{0.4pt}}
  \vspace{8pt}
  \begin{center}
  \begin{minipage}{0.92\textwidth}
  \small
  \noindent\textbf{Abstract.}\quad
Modeling social media public opinion evolution is essential for governance decision-making. Traditional epidemic models and rule-based agent-based models (ABMs) fail to capture the cognitive processes and adaptive behaviors of real users. Recent large language model (LLM)-based social simulations can reproduce group-level phenomena like polarization and conformity, yet remain unable to recreate the irrational interactions and multi-phase dynamics of real public opinion events. We present POSIM (\textbf{P}ublic \textbf{O}pinion \textbf{Sim}ulator), a multi-agent simulation framework for social media public opinion evolution and governance. POSIM integrates LLM-driven agents with a Belief--Desire--Intention (BDI) cognitive architecture that accounts for irrational factors, places them in a virtual social media environment with social networks and recommendation mechanisms, and drives temporal dynamics through a Hawkes point process engine that captures the co-evolution of agents and the environment across event phases. To validate the framework, we collect real-world public opinion datasets from the Weibo platform covering the full interaction chain of users. Experiments show that POSIM successfully reproduces key characteristics of public opinion evolution from individual mechanisms to collective phenomena, and its effectiveness is further supported by multiple statistical metrics. Building on POSIM, governance-oriented guidance and intervention experiments uncover a counterintuitive \textit{empathy paradox}: empathetic guidance deepens negative sentiment instead of easing it under certain conditions, offering new insights for governance strategy design. These results demonstrate that the proposed framework can fully serve as a computational experimentation platform for proactive strategy evaluation and evidence-based governance. 
All source code is available at \url{https://github.com/DeepCogLab/posim/}.
  \end{minipage}
  \end{center}
  \vspace{6pt}

  \begin{center}
  \begin{minipage}{0.92\textwidth}
  \end{minipage}
  \end{center}
  \vspace{4pt}
  \centerline{\rule{0.92\textwidth}{0.4pt}}
  \vspace{14pt}
]

\section{Introduction}\label{sec:intro}

Social media has become the primary arena where public opinion forms, spreads, and evolves. A single breaking event can trigger mass participation within hours, accompanied by collective emotional arousal~\citep{gonzalez_emotional_2010} and information cascades~\citep{goldstein_structure_2012}. Understanding and anticipating such dynamics is therefore critical for social governance~\citep{WU2025127823}. Yet real-world social experiments are constrained by ethical concerns, irreproducibility, and uncontrollable variables, making it impractical to repeatedly test mechanisms and intervention strategies in live settings. Computational simulation has thus become a key methodology for studying public opinion evolution and informing governance decisions~\citep{bonabeau_agent-based_2002}.

Building a simulation that faithfully recreates real public opinion dynamics poses several challenges. Social media users do not make purely rational decisions; their behaviors are shaped by the interplay of emotions, beliefs, and cognitive biases~\citep{buckels2014trolls}, and a credible simulation must capture this cognitive complexity. Moreover, public opinion evolution shows pronounced temporal dynamics, as events can rapidly shift through outbreak, decay, and resurgence phases~\citep{ZHOU2024121307}, placing strict demands on temporal modeling~\citep{zhao_seismic_2015}. Beyond behavioral fidelity, any simulation must address the question of why its results should be trusted: it should not only match real data at the aggregate level but also demonstrate that its underlying mechanisms are plausible and interpretable~\citep{mcculloch_calibrating_2022}. Traditional epidemic models~\citep{daley_stochastic_1965} and threshold cascade models~\citep{granovetter_threshold_1978} reduce individuals to a few discrete states, unable to express rich semantics, interaction strategies, or user heterogeneity. Agent-based modeling (ABM) offers a bottom-up paradigm~\citep{bonabeau_agent-based_2002} often combined with opinion dynamics rules~\citep{castellano_statistical_2009}, yet its rule-driven agents still cannot capture the cognitive processes, nuanced interactions, and environment coupling of real users.

Large language model (LLM)-driven social simulation offers a promising alternative~\citep{wang_survey_2024}. Generative Agent-Based Modeling (GABM)~\citep{park_generative_2023} has shown that LLM-powered agents can produce human-like social behaviors and give rise to emergent phenomena such as polarization, conformity, and echo chambers~\citep{gao_s3_2023,chuang_simulating_2024}. However, existing GABM frameworks face three key limitations when applied to real public opinion evolution. First, most treat the LLM as an end-to-end behavior generator without explicitly modeling the cognitive pipeline from belief formation through motivation to decision-making, leaving them unable to represent how irrational factors such as emotional arousal and cognitive biases shape user decisions. Second, existing work largely targets abstract opinion scenarios and has not effectively captured the co-evolutionary dynamics linking environmental triggers, user responses, and propagation amplification. Third, validation efforts lack systematic rigor, leaving the credibility of simulation outputs poorly established. A framework that targets real public opinion evolution while ensuring result credibility therefore remains absent.

To address these limitations, we propose POSIM (\textbf{P}ublic \textbf{O}pinion \textbf{Sim}ulator), a multi-agent simulation framework for social media public opinion evolution and governance. POSIM constructs LLM-powered agents grounded in Belief--Desire--Intention (BDI) cognitive theory~\citep{bratman_intention_1987,rao_bdi_1995}, incorporating emotional arousal and cognitive biases into the BDI architecture to equip agents with explicit cognitive states and human-like irrational behaviors. These agents operate within a virtual social media environment with social networks and recommendation mechanisms, supporting realistic opinion behaviors such as browsing, expression, and reposting. A temporal dynamics engine based on the Hawkes self-exciting point process~\citep{hawkes_spectra_1971} unifies exogenous event shocks and endogenous user interactions to drive multi-phase opinion evolution.

Validating the effectiveness of a public opinion simulation system requires more than statistical consistency with real data; the internal mechanisms must also be plausible and interpretable, avoiding what we term \textit{superficial agreement}, where outputs match but processes remain opaque. To credibly validate POSIM's simulation fidelity, we establish a three-tier progressive validation framework spanning mechanism, phenomenon, and statistics, respectively addressing: (1)~Are the internal mechanisms sound? (2)~Are the emergent group phenomena plausible? (3)~Are the statistical outputs reliable? We collect three real-world public opinion datasets from the Weibo platform covering the full interaction chain of original posts, reposts, and comments, and conduct systematic experiments. Results confirm that POSIM exhibits sound agent behavioral logic at the mechanism tier, reproduces group phenomena consistent with social science theory at the phenomenon tier, and achieves close alignment with real opinion data at the statistical tier. Moreover, governance-oriented experiments reveal that seemingly positive strategies can produce counterproductive effects in complex opinion interactions, providing important empirical evidence for governance strategy design. The main contributions are:

\begin{itemize}[leftmargin=1.5em, labelsep=0.5em, label={\textbullet}]
\item We propose POSIM, a multi-agent simulation framework for social media public opinion evolution and governance, advancing the modeling objective from abstract opinion scenarios to credible simulation of real public opinion events.
\item We collect three real-world public opinion datasets from Weibo spanning the full interaction chain, and establish a mechanism--phenomenon--statistics three-tier progressive validation framework that calibrates from agent decision quality through emergent group phenomena to multi-dimensional statistical metrics, providing a reproducible credibility evaluation paradigm for LLM-driven social simulation.
\item Governance-oriented experiments on POSIM reveal an \textit{empathy paradox}: empathetic guidance intended to ease group antagonism instead amplifies negative sentiment in complex multi-agent interactions, providing critical empirical evidence and risk insights for cognitive-level governance strategy design.
\end{itemize}

\section{Related Work}\label{sec:related}

\subsection{Public Opinion Propagation Modeling}
Public opinion propagation modeling studies how information spreads and evolves in social networks. Early work borrowed from epidemiology: the stochastic rumor model~\citep{daley_stochastic_1965} classifies individuals into ignorant, spreader, and stifler states. Later studies incorporated complex network topologies~\citep{watts_strogatz_1998,barabasi_albert_1999} and multi-stage transition mechanisms~\citep{jiang_weibo_spnr_2020} to examine how network structure shapes cascade scale, and threshold cascade models~\citep{granovetter_threshold_1978,watts_cascades_2002} explain collective behavior as emerging from individual decision thresholds. System dynamics methods take a different approach, modeling causal relationships among actors through stock-flow diagrams and feedback loops~\citep{gao_sd_public_opinion_2019,liu_evolution_2024,CHENG2026103484}. These methods capture macroscopic opinion trajectories but treat populations as continuous aggregate flows, overlooking individual heterogeneity. Self-exciting point processes~\citep{hawkes_spectra_1971} and their variants offer continuous-time tools for modeling social media cascades and have been widely used for popularity prediction, yet they too focus on population-level statistical regularities rather than individual cognition. In summary, these approaches each have strengths in modeling collective trends, but none explicitly addresses individual cognitive processes, motivating the shift toward agent-based paradigms.

\subsection{Agent-Based Modeling and Opinion Dynamics}

Agent-based modeling (ABM) provides a bottom-up paradigm for studying social systems: heterogeneous individuals follow local behavioral rules, and collective phenomena emerge from their interactions~\citep{bonabeau_agent-based_2002}. Parameter calibration methods~\citep{mcculloch_calibrating_2022} further enable systematic validation. ABM has produced a rich body of work in opinion dynamics. The DeGroot model~\citep{degroot_reaching_1974} describes consensus through weighted averaging; the Deffuant--Weisbuch model~\citep{deffuant_mixing_2000} introduces pairwise bounded interactions; and the Hegselmann--Krause model~\citep{hegselmann_opinion_2002} explains group fragmentation through selective exposure. Together, these models reproduce phase transitions from consensus to polarization~\citep{castellano_statistical_2009,bernardo_bc_survey_2024}. More recent work has built multi-type opinion simulation systems on social networks~\citep{chen_evolution_2025,wu_public_2023,LI2026103965} to study polarization and echo chambers~\citep{liu_agent-based_2024}. However, traditional ABM agents rely on predefined numerical rules: opinions are reduced to low-dimensional continuous variables and interactions are limited to weighted averages or threshold functions. Such rule-driven agents cannot perceive the rich information in real opinion environments, nor can they simulate cognitive processes such as emotional evolution, motivational reasoning, and autonomous decision-making.

\subsection{LLM-Driven Social Simulation}

Large language models (LLMs), with their capacity for semantic understanding, contextual reasoning, and autonomous decision-making, have opened new possibilities for social simulation beyond rule-based paradigms~\citep{wang_survey_2024,WU2026103791}. Generative Agents~\citep{park_generative_2023} first demonstrated that LLM-powered agents can autonomously plan and engage in social interactions, establishing the GABM paradigm. GABMs have since been applied to governance-related domains~\citep{repec:jas:jasssj:2025-107-3}, though a reliable system for simulating complex public opinion evolution remains absent.

Research on LLM-driven opinion simulation has expanded rapidly. S3~\citep{gao_s3_2023} was among the first to build an LLM-based information propagation framework; GA-S3~\citep{zhang_ga_s3_2025} introduced group agents for multi-layered network interactions but lacked deep bottom-up emergence; SPARK~\citep{zhang_spark_2025} and LLM-AIDSim~\citep{zhang_llm-aidsim_2025,hu_llm-enhanced_2024} explored stance evolution and information diffusion patterns; FDE-LLM~\citep{yao_social_2025} fused dynamical equations with language models to impose mathematical constraints on opinion trajectories. On the scalability front, OASIS~\citep{yang_oasis_2025} scaled to the million-node level; LMAgent~\citep{liu_lmagent_2026} built a multi-modal social simulation system with limited cross-scenario transferability; HiSim~\citep{mou_unveiling_2024} adopted a hybrid strategy balancing efficiency and fidelity but could not reproduce multi-phase hotness evolution; TrendSim~\citep{zhang_trendsim_2025} simulated trending topic interactions but lacked systematic validation. As summarized in Table~\ref{tab:comparison}, existing frameworks generally treat LLMs as end-to-end behavior generators without explicitly modeling the cognitive pathway from belief formation through motivational reasoning to behavioral output. They also lack integration of environmental dynamics and temporal precision, and their validation remains fragmented, making them insufficient for credible governance-oriented evaluation.

To address these gaps, POSIM introduces a systematic design spanning agents, environment, and strategy evaluation, providing an integrated solution for credible public opinion simulation and governance decision support.

\begin{table}[!t]
\centering
\caption{Comparison of POSIM with representative public opinion simulation frameworks.}
\label{tab:comparison}
\renewcommand{\arraystretch}{1.15}
\resizebox{\columnwidth}{!}{%
\scriptsize
\begin{tabular}{@{}l c ccc c c cc@{}}
\toprule
\textbf{Platform} &
\makecell{\textbf{Explicit}\\\textbf{Cognitive}\\\textbf{Modeling}} &
\multicolumn{3}{c}{\textbf{Validation Ladder}} &
\makecell{\textbf{Real-Case}\\\textbf{Intervention}} &
\makecell{\textbf{LLM-Based}\\\textbf{Multi-Type}\\\textbf{Agents}} &
\makecell{\textbf{Temporal}\\\textbf{Precision}} &
\makecell{\textbf{Modular}\\\textbf{Design}} \\
\cmidrule(lr){3-5}
 & & \textbf{M} & \textbf{P} & \textbf{S} & & & & \\
\midrule
S3~\citep{gao_s3_2023}
  & $\times$ & $\times$ & $\checkmark$ & $\checkmark$
  & $\times$ & $\times$
  & $\bigstar\bigstar\bigstar$ & $\bigstar\bigstar\bigstar$ \\
HiSim~\citep{mou_unveiling_2024}
  & $\times$ & $\times$ & $\times$ & $\checkmark$
  & $\times$ & $\times$
  & $\bigstar\bigstar$ & $\bigstar\bigstar$ \\
GA-S3~\citep{zhang_ga_s3_2025}
  & $\times$ & $\times$ & $\times$ & $\checkmark$
  & $\times$ & $\checkmark$
  & $\bigstar\bigstar\bigstar$ & $\bigstar\bigstar$ \\
SPARK~\citep{zhang_spark_2025}
  & $\times$ & $\times$ & $\checkmark$ & $\times$
  & $\times$ & $\checkmark$
  & $\bigstar\bigstar$ & $\bigstar\bigstar$ \\
FDE-LLM~\citep{yao_social_2025}
  & $\times$ & $\times$ & $\times$ & $\checkmark$
  & $\times$ & $\times$
  & $\bigstar\bigstar$ & $\bigstar\bigstar$ \\
TrendSim~\citep{zhang_trendsim_2025}
  & $\times$ & $\checkmark$ & $\times$ & $\times$
  & $\times$ & $\checkmark$
  & $\bigstar\bigstar\bigstar\bigstar$ & $\bigstar\bigstar\bigstar$ \\
OASIS~\citep{yang_oasis_2025}
  & $\times$ & $\times$ & $\checkmark$ & $\checkmark$
  & $\times$ & $\times$
  & $\bigstar\bigstar\bigstar\bigstar$ & $\bigstar\bigstar\bigstar\bigstar$ \\
LMAgent~\citep{liu_lmagent_2026}
  & $\times$ & $\times$ & $\times$ & $\checkmark$
  & $\times$ & $\checkmark$
  & $\bigstar\bigstar$ & $\bigstar\bigstar$ \\
\midrule
\rowcolor{grayrow}
\textit{\textbf{POSIM (Ours)}}
  & $\checkmark$ & $\checkmark$ & $\checkmark$ & $\checkmark$
  & $\checkmark$ & $\checkmark$
  & $\bigstar\bigstar\bigstar\bigstar\bigstar$ & $\bigstar\bigstar\bigstar\bigstar\bigstar$ \\
\bottomrule
\end{tabular}%
}
\par\vspace{2pt}
{\scriptsize\rmfamily
\parbox{\columnwidth}{\raggedright \textit{Note}: $\bigstar$ denotes capability level (1--5); M, P, and S denote mechanism, phenomenon, and statistical validation; real-case intervention denotes governance-oriented intervention experiments conducted on real public opinion events, beyond merely using real data for validation.}}
\end{table}

\section{Method}\label{sec:method}

\begin{figure*}[!t]
\centering
\includegraphics[width=0.88\textwidth]{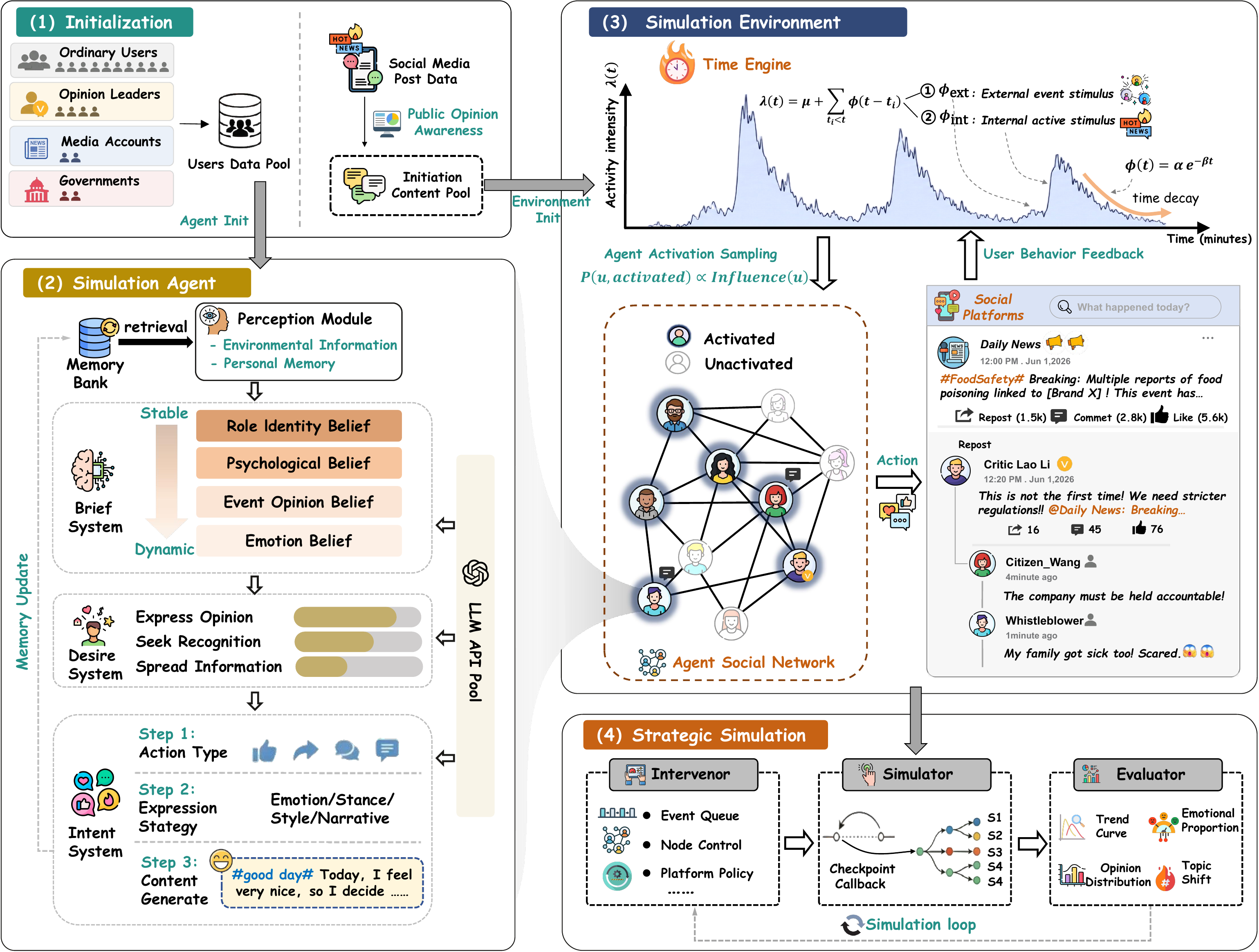}
\caption{Overall architecture of POSIM. (1)~The initialization module maps real user data and post data to agent profiles and environment initial states. (2)~The Social-BDI agent architecture implements the cognitive pipeline: Perception~$\to$~Belief~$\to$~Desire~$\to$~Intention~$\to$~Action. (3)~The simulation environment consists of a virtual social media platform and a temporal engine. The platform provides personalized content recommendations, and the temporal engine governs agent activation rhythms based on the Hawkes self-exciting point process. Agent behavioral outputs feed back into the environment to drive subsequent activations, forming a continuous interaction loop. (4)~Strategy evaluation uses this loop as its simulator core, with an intervenor for injecting governance strategies and an evaluator for quantitative analysis, supporting counterfactual assessment.}
\label{fig:framework}
\end{figure*}

\subsection{Framework Overview}\label{subsec:overview}

As shown in Figure~\ref{fig:framework}, POSIM comprises three core components addressing who participates, in what environment they interact, and how simulation supports decision-making. After initialization (Figure~\ref{fig:framework}(1)), agents and environment form a continuous interaction loop that strategy evaluation leverages for counterfactual governance assessment.

At the agent level (Figure~\ref{fig:framework}(2); Section~\ref{subsec:agent}), the Social-BDI architecture decomposes cognition into three LLM-powered subsystems: belief update maintains four hierarchical cognitive layers (role identity, psychological cognition, event opinion, and emotional arousal); desire generation maps beliefs to behavioral motivations; and intention planning translates motivations into concrete actions via multi-level chain-of-thought reasoning. Four agent types (ordinary users, opinion leaders, media accounts, and governments) share this pipeline but produce differentiated behaviors through role-specific prompts.

At the environment level (Figure~\ref{fig:framework}(3); Section~\ref{subsec:env}), the Hawkes-based temporal engine enables minute-level resolution and naturally reproduces multi-phase evolution patterns. Agent outputs feed back as endogenous excitation, forming a bidirectional interaction loop with the platform (Section~\ref{subsec:interaction}).

At the strategy evaluation level (Figure~\ref{fig:framework}(4); Section~\ref{subsec:strategy}), checkpoint callback enables counterfactual comparison by branching from identical states under different interventions.

The following sections detail the design of each component.

\subsection{Social-BDI Agent Architecture}\label{subsec:agent}

Traditional reactive agents follow a simple perception-to-action mapping and cannot capture the cognitive mechanisms behind social media user decisions. In practice, users' decisions are shaped by the interplay of emotions, cognitive biases, and social influence, and the coupling of emotion and intention is a key driver of crisis event evolution~\citep{myint2024crisis_mtl_intent}. Building on the classic BDI architecture~\citep{bratman_intention_1987,rao_bdi_1995}, we propose Social-BDI, which integrates emotional arousal and cognitive biases into the cognitive framework to better capture the irrational characteristics of real social media users. Social-BDI has two distinguishing features. First, agents maintain explicit cognitive states, with every decision grounded in observable state changes. Second, the cognitive process forms an auditable multi-stage decision chain~\citep{wei_chain_2022} rather than a single black-box call. The pipeline is formalized as:
\begin{equation}
\underset{\text{Perception}}{P_t} \longrightarrow \underset{\text{Belief}}{B_t} \longrightarrow \underset{\text{Desire}}{D_t} \longrightarrow \underset{\text{Intention}}{I_t} \longrightarrow \underset{\text{Action}}{A_t}
\end{equation}

The architecture uses an LLM as its reasoning engine and comprises three modules: belief, desire, and intention. These subsystems execute sequentially in each decision cycle, with each receiving upstream output as input, forming a traceable pipeline from perception to action.

\subsubsection{Belief Subsystem}

Beliefs represent an agent's cognitive model of the external world and its internal state, forming the basis for all reasoning and decisions. Different cognitive layers vary in stability: core personality traits remain stable in the short term, while event-specific opinions and emotions shift rapidly with new information. Based on this stratification, we design a hierarchical belief system $B = \langle B^{id}, B^{psy}, B^{evt}, B^{emo} \rangle$, ordered from most stable to most volatile.

\textbf{Role identity belief}~$B^{id}$ encodes an agent's foundational attributes, including demographics (gender, region), social properties (follower count, verification type), and identity traits (description, interest tags). During initialization, user information and historical posts from the real platform are processed by an LLM to generate a first-person role identity description. This belief remains fixed throughout simulation, anchoring personality consistency.

\textbf{Psychological cognition belief}~$B^{psy}$ captures the psychological characteristics and cognitive tendencies an agent exhibits in public opinion participation. Users display typical patterns such as conformity, prejudice, impulsiveness, and obstinacy when engaging with online events~\citep{buckels2014trolls}. Based on behavioral analysis of real events, we define representative psychological types: self-actualization, curiosity-seeking, cathartic venting, anti-authority resentment, and bandwagon conformity. Each type carries predefined cognitive entries; for example, curiosity-seeking includes ``there must be undisclosed details behind this event'' and ``the more information is suppressed, the more it deserves investigation,'' while cathartic venting includes ``real-life frustrations can be vented online'' and ``the harsher the words, the more satisfying it feels.'' During initialization, an LLM samples the best-fitting profile from a predefined trait pool based on each user's historical behavior. These beliefs remain highly stable throughout simulation, reflecting the persistence of deep psychological traits.

\textbf{Event opinion belief}~$B^{evt}$ represents an agent's factual understanding, contextual awareness, and stance toward specific actors in an event. Unlike the preceding layers, event opinions evolve dynamically as new information or viewpoints trigger revision or reinforcement. Each opinion is stored as a structured quadruple $\langle t, \text{subject}, \text{opinion}, \text{reason} \rangle$. Initial stances are extracted through LLM-driven structured interviews based on the user's pre-event posts. For example, the LLM is prompted with questions such as ``What is your stance toward the involved party in event X'' to elicit structured stance descriptions from historical behavior.

\textbf{Emotional arousal belief}~$B^{emo}$ draws on the ABC model of cognitive behavioral theory~\citep{ellis_reason_1962}, which highlights cognitive appraisal as a mediator of emotional responses. We adopt Ekman's six basic emotions~\citep{ekman_basic_1992}, representing emotional states as:
\begin{equation}
\mathbf{e} = [e_{\text{hap}},\, e_{\text{sad}},\, e_{\text{ang}},\, e_{\text{fear}},\, e_{\text{sur}},\, e_{\text{dis}}]^\top \in [0,1]^6
\end{equation}
Emotions are updated by three mechanisms. \textbf{Temporal decay} reduces intensity exponentially:
\begin{equation}
\mathbf{e}_i(t) = \mathbf{e}_i(t_0) \cdot e^{-\lambda_e(t-t_0)}
\end{equation}
\textbf{Content stimulation} adds the impact of recommended content: $\mathbf{e}_i \leftarrow \mathbf{e}_i + \eta \cdot \mathbf{s}_{content}$, where $\eta$ is the stimulation coefficient and $\mathbf{s}_{content}$ is the stimulus vector from LLM sentiment analysis. \textbf{Social contagion}, based on emotional contagion theory~\citep{hatfield_emotional_1994}, propagates neighboring emotions:
\begin{equation}
\mathbf{e}_i \leftarrow (1-\rho) \cdot \mathbf{e}_i + \rho \cdot \bar{\mathbf{e}}_{\text{neighbor}}
\end{equation}
where $\rho$ is the contagion coefficient and $\bar{\mathbf{e}}_{\text{neighbor}}$ is the weighted average of neighbors' emotions.

\textbf{Belief update mechanism.} At each time step, the LLM updates beliefs based on two inputs: environmental perception $P_t$ and historical memory $M_t$. Memories are stored in a streaming memory bank and retrieved by a scoring function that balances recency and semantic relevance:
\begin{equation}
R(m, t) = \alpha_{\mathrm{rec}} e^{-\gamma(t - t_m)} + (1-\alpha_{\mathrm{rec}})\,\text{sim}(q, m),
\label{eq:mmr}
\end{equation}
where $\text{sim}(\cdot,\cdot)$ is cosine similarity from a pretrained encoder and $\alpha_{\mathrm{rec}}\in[0,1]$ controls the recency--relevance trade-off. In practice, users' belief updates are far from rational; they are systematically shaped by cognitive biases~\citep{tversky_judgment_1974}. Users selectively accept information consistent with existing views (\textit{confirmation bias}) and under high arousal make emotion-driven judgments bypassing rational analysis (\textit{affect-driven reasoning}). For ordinary users, we therefore introduce explicit cognitive bias guidance into the LLM reasoning via prompts (Appendix~\ref{app:prompts}). The overall update is:
\begin{equation}
B_{t} = f_{\text{LLM}}(B_{t-1}, P_t, M_t, \Phi)
\label{eq:belief_update}
\end{equation}
where $\Phi$ denotes cognitive bias guidance embedded in prompts. Here $B_{t-1}$ is the belief state carried over from the previous step. The update primarily affects the fast-changing layers $B^{evt}$ and $B^{emo}$, while $B^{psy}$ evolves slowly; $B^{id}$ remains fixed as the identity anchor.

\subsubsection{Desire Subsystem}

The desire subsystem bridges cognitive states and behavioral motivations, connecting ``how the agent understands the world'' with ``what the agent wants to do.'' In classical BDI theory, desires are goal states the agent wishes to achieve~\citep{bratman_intention_1987}; in social media, we operationalize them as intrinsic motivations for participating in opinion discourse. Motivations are highly diverse: the same news may trigger a sense of justice in one user and curiosity in another. This heterogeneity cannot be captured by fixed rules, so we use the LLM's commonsense reasoning to infer motivations from belief states.

At each time step, the desire subsystem receives the updated beliefs $B_{t}$ and current perception $P_t$, outputting motivation descriptions with intensity labels:
\begin{equation}
\begin{split}
D_t &= \{(d_1, l_1), (d_2, l_2), \ldots, (d_n, l_n)\}, \\
l_i &\in \{\text{very low},\; \text{low},\; \text{medium},\; \text{high},\; \text{very high}\}
\end{split}
\end{equation}
where $d_i$ is a natural-language motivation and $l_i$ is a five-level intensity label. Uses and Gratifications theory~\citep{katz_uses_1973} posits that audiences actively select media to fulfill psychological needs. Inspired by this, we predefine candidate motivation sets per agent type, from which the LLM dynamically selects and weights based on belief state. Ordinary users' candidates include emotional venting, justice advocacy, social connection, etc.; opinion leaders additionally include self-expression, influencing others, etc. The combination and intensity are context-dependent: high anger in $B^{emo}$ with a strong negative stance in $B^{evt}$ elevates emotional venting; a curiosity-dominated $B^{psy}$ foregrounds information acquisition.

The motivation list produced here will be passed to the downstream intention subsystem (described next), where high-weight motivations constrain the selection of action types and expression strategies. For instance, dominant emotional venting tends to result in short, emotionally charged comments, while dominant self-expression tends to result in lengthy original posts with rational analysis. The desire generation function is:
\begin{equation}
D_t = g_{\text{LLM}}(B_{t}, P_t)
\end{equation}

The key contribution is an explicit motivational layer between beliefs and actions: beliefs first produce structured motivations, which the intention subsystem then translates into concrete behaviors. The effect of this design on generating realistic discourse is validated in the ablation study (Section~\ref{sec:exp}).

\subsubsection{Intention Subsystem}

The intention subsystem serves as the final decision stage, functioning as a hierarchical behavior planner. Following social media interaction conventions, we define atomic behavioral categories: like, direct repost, repost with comment, short comment, long comment, short original post, and long original post. The subsystem employs a multi-level chain-of-thought mechanism that decomposes decisions into three progressive levels.

\textbf{Level 1: Action type and target selection.} The LLM selects an action type based on belief state and desire intensity, and determines the interaction target. This level decides what to do and to whom.

\textbf{Level 2: Expression strategy planning.} When generating text, the system plans strategies along four dimensions: emotion (type and intensity), stance (support, opposition, or neutrality), style (e.g., rational analysis, sarcasm, emotional aggression), and narrative (e.g., factual enumeration, labeling, call to action). The LLM selects the most fitting combination. This level decides how to express.

\textbf{Level 3: Content generation.} Constrained by prior decisions, the system generates concrete text matching the agent's role, including body text, hashtags, and mentions. This level decides what exactly to say.

This three-level decomposition provides explicit constraints that prevent the LLM from producing bland, averaged text. Every decision is logged for full traceability. The intention generation is:
\begin{equation}
I_t = h_{\text{LLM}}(B_{t}, D_t, P_t)
\end{equation}
where $I_t = \{(a_j, o_j, s_j, c_j)\}_{j=1}^{m}$ is the action sequence, each element comprising action type $a_j$, target $o_j$, strategy vector $s_j = (\text{emotion}, \text{stance}, \text{style}, \text{narrative})$, and content $c_j$.

\subsubsection{Heterogeneous Agents}

Real-world public opinion evolution involves interactions among ordinary users, opinion leaders, media accounts, and governments~\citep{chen_evolution_2025,wu_public_2023}. All four types share the unified Social-BDI architecture. Behavioral differences arise solely from type-specific role-guiding prompts that inject distinct role characteristics (identity, speaking style, focal concerns), enabling the same architecture to produce differentiated behaviors. This ensures that heterogeneity stems from role differences rather than architectural modifications, preserving both unity and interpretability.

\textbf{Ordinary users} form the majority of participants, with behaviors primarily shaped by personal beliefs and immediate emotions. Their language is colloquial and fragmented, and high arousal readily produces impulsive expression.

\textbf{Opinion leaders} serve as critical information intermediaries according to two-step flow theory~\citep{katz_personal_1955,choi2015two}, with large follower bases and significant social influence. Their outputs tend toward independence and agenda-setting, guiding downstream users' belief updates and decisions.

\textbf{Media accounts} prioritize timeliness, accuracy, and authoritativeness. Their language is formal and restrained, participating primarily through news reports, updates, and in-depth commentary, serving to confirm information and frame agendas at key junctures.

\textbf{Governments} represent official positions and governance perspectives, posting infrequently but with high authority, typically issuing statements after events reach sufficient scale, exerting pivotal influence on opinion trajectories.

All four types generate their behavioral patterns through the Social-BDI pipeline, constrained by their respective role-guiding prompts.

Complete cognitive pipeline examples appear in Appendix~\ref{app:prompts}. Pseudocode for the Social-BDI pipeline is given in Algorithm~\ref{alg:Social-BDI}.

\begin{algorithm}[!t]
\footnotesize
\caption{Social-BDI Cognitive Decision Pipeline}
\label{alg:Social-BDI}
\begin{algorithmic}[1]
\REQUIRE Agent $u$, perception $P_t$, event context $\mathcal{E}_{ctx}$
\ENSURE Action sequence $I_t$
\STATE \textbf{Phase 1: Belief Update}
\STATE $M_t \leftarrow \text{Memory.get}(P_t, k)$
\STATE $B_t \leftarrow f_{\text{LLM}}(B_{t-1}, P_t, M_t, \Phi)$
\STATE $B_t.\text{emo} \leftarrow B_t.\text{emo} \odot e^{-\lambda_e \Delta t}$
\STATE \textbf{Phase 2: Desire Generation}
\STATE $D_t \leftarrow g_{\text{LLM}}(B_t, P_t)$
\STATE \textbf{Phase 3: Intention Planning}
\STATE $L_1 \leftarrow \text{SelectAction}(B_t, D_t, P_t)$
\STATE $L_2 \leftarrow \text{PlanStrategy}(L_1, B_t)$
\STATE $L_3 \leftarrow \text{GenContent}(L_1, L_2, B_t)$
\STATE $I_t \leftarrow (L_1, L_2, L_3)$
\STATE \textbf{Phase 4: Memory Update}
\STATE $\text{Memory.store}(I_t, t)$
\RETURN $B_t,\, D_t,\, I_t$
\end{algorithmic}
\end{algorithm}

\subsection{Simulation Environment}\label{subsec:env}

The simulation environment provides a virtual space for agent perception and interaction, designed to reproduce the information flow and temporal dynamics of social media while remaining computationally tractable. As shown in Figure~\ref{fig:framework}, it comprises two components: the \textbf{virtual social media platform}, which determines what agents see, and the \textbf{temporal engine}, which determines when and which agents are activated. Together they drive continuous agent--environment interaction.

\subsubsection{Virtual Social Media Platform}

The platform constructs a personalized information environment for each activated agent, determining what content they see and with whom they interact.

\textbf{Agent social network.} POSIM maintains three directed networks: a follower network, a repost network, and a comment network. The follower network records follow relationships and defines the scope of relationship-based recommendations. The repost and comment networks grow dynamically as each interaction updates the topology.

\textbf{Content recommendation.} The system maintains a global content pool updated in real time, with vector representations from a pretrained semantic encoder (BGE-small-zh\footnote{\url{https://huggingface.co/BAAI/bge-small-zh-v1.5}}) for cosine-based deduplication. Content candidates come from two channels: the \textbf{relationship channel} retrieves posts from followed users, and the \textbf{public channel} retrieves from the global pool, simulating cross-circle exposure. Candidates are scored along three dimensions. \textbf{Homophily}~$H(u,p)$ measures cosine similarity between user and content embedding vectors. \textbf{Popularity}~$V(p)$ uses normalized engagement signals (likes, reposts, comments). \textbf{Freshness}~$F(p)$ applies exponential time decay, favoring recent content. The final score is:
\begin{equation}
S_{exp}(u, p) = w_h \cdot H(u, p) + w_p \cdot V(p) + w_r \cdot F(p)
\label{eq:recommend}
\end{equation}
where $w_h$, $w_p$, $w_r$ are adjustable weights (Appendix Table~\ref{tab:rec_params}). After ranking, $r_{explore}=0.2$ of slots are reserved for random exploration, and a history mechanism records each agent's previously viewed content to prevent repeated exposure within a sliding window. The system also tracks hashtag popularity and periodically injects trending topics into recommendations, simulating the attention-focusing effect of real trending lists.

\subsubsection{Temporal Engine}

User activity in real public opinion events exhibits highly non-uniform temporal distributions: breaking news can trigger thousands of reposts within minutes, while activity drops sharply during lulls. Conventional approaches activate agents at fixed intervals independent of external stimuli, where each agent's activation probability is identical and uncorrelated with ongoing events. Such independent uniform activation fails to reproduce these event-driven surges.

POSIM adopts a \textbf{temporal engine} based on the Hawkes self-exciting point process as its temporal scheduler. The key insight is that collective activity is not a preset constant but a dynamic quantity driven by exogenous shocks and endogenous interactions. Each event briefly elevates the probability of subsequent events, much like a viral post stimulating further participation before the effect decays. The conditional intensity is:
\begin{equation}
\lambda(t) = \mu + \sum_{t_i < t} \phi(t - t_i)
\label{eq:hawkes}
\end{equation}
where $\mu$ is the background rate reflecting baseline activity, and $\phi(t) = \alpha e^{-\beta t}$ is the exponentially decaying excitation kernel. We decompose excitation into \textbf{exogenous}~$\phi_{ext}$ and \textbf{endogenous}~$\phi_{int}$ components (Figure~\ref{fig:framework}):
\begin{equation}
\phi_{ext}(t) = \alpha_{ext} \cdot e^{-\beta_{ext} t}, \quad \phi_{int}(t) = \alpha_{int} \cdot e^{-\beta_{int} t}
\label{eq:excitation}
\end{equation}
with $\alpha_{ext} \gg \alpha_{int}$ and $\beta_{ext} < \beta_{int}$. Exogenous excitation corresponds to critical external stimuli (e.g., breaking news, official statements) with strong intensity and slow decay. These are triggered by an \textbf{event queue}, a preconfigured timeline injected into the Hawkes process at designated times. Endogenous excitation corresponds to user interactions with weaker intensity and faster decay, reflecting the self-exciting nature of social propagation. Their superposition naturally produces ``outbreak--sustain--decay'' patterns consistent with real public opinion events.

We further incorporate a circadian rhythm modulation mechanism, scaling Hawkes intensity with a 24-hour periodic function. At each step~$\Delta t$, the number of activated agents is determined by Poisson sampling
\begin{equation}
N_t \sim \text{Poisson}(\lambda(t) \cdot N_{s} \cdot \Delta t \,/\, 60)
\end{equation}
with specific agents selected via activity-weighted sampling. Full parameter configuration is in Appendix~\ref{app:params}; computational resource consumption is reported in Appendix~\ref{app:efficiency}.

\subsection{Agent--Environment Interaction}\label{subsec:interaction}

At each step, the temporal engine determines the activated agent set $\mathcal{A}_t$ based on Hawkes intensity, and the recommendation system generates personalized feeds for each agent. These feeds, together with any event triggers, are merged by the perception module into environmental perception $P_t$, which drives the Social-BDI pipeline to produce behaviors $A_t$. The resulting outputs then flow back into the environment, where new content enters the global pool, interactions update network topology, and activity contributes endogenous excitation to the Hawkes process, forming a continuous bidirectional loop.

This loop mirrors real social media information flow. Along the forward path, the recommendation system scores content by homophily, popularity, and freshness (Eq.~\ref{eq:recommend}), shaping each agent's information environment. Along the reverse path, behavioral outputs introduce new content and engagement signals that reshape subsequent recommendations, closing the feedback cycle of information shaping cognition, cognition driving behavior, and behavior reshaping information. The Hawkes process further amplifies this dynamic, as dense behaviors triggered by exogenous events sustain elevated activity through endogenous excitation, producing nonlinear patterns consistent with real events. Consequently, collective phenomena such as opinion lifecycle, emotional polarization, and information cascades emerge from individual interactions, while each agent's behavior remains grounded in its cognitive process. Pseudocode is given in Algorithm~\ref{alg:main}.

\begin{algorithm}[!t]
\footnotesize
\caption{POSIM Main Simulation Loop}
\label{alg:main}
\begin{algorithmic}[1]
\REQUIRE Agent set $\mathcal{U}$, event queue $\mathcal{E}$, step size $\Delta t$, total steps $T_{max}$
\ENSURE Simulation trajectory $\mathcal{T}$
\STATE Initialize Hawkes time engine ($\mu,\, \phi_{ext},\, \phi_{int}$) and recommendation system
\FOR{$t = 0, 1, \ldots, T_{max}$}
 \FORALL{exogenous event $e \in \mathcal{E}$ triggered at step $t$}
 \STATE Inject $e$ into Hawkes process as exogenous excitation $\phi_{ext}$
 \ENDFOR
 \STATE Compute intensity $\lambda(t)$ (Eq.~\ref{eq:hawkes})
 \STATE $N_t \sim \text{Poisson}(\lambda(t) \cdot N_s \cdot \Delta t\, /\, 60)$
 \STATE Sample activated set $\mathcal{A}_t \subset \mathcal{U}$, $|\mathcal{A}_t|=N_t$, weighted by historical activity
 \FORALL{$u \in \mathcal{A}_t$ \textbf{parallel}}
 \STATE Generate personalized feed for $u$; merge with exogenous events as perception $P_t$
 \STATE $B_{t},\, D_t,\, I_t \leftarrow \text{Social-BDI}(u,\, P_t)$ \COMMENT{Alg.~\ref{alg:Social-BDI}}
 \STATE $A_t \leftarrow \text{Parse}(I_t)$; execute $A_t$; update content pool and social network
 \STATE Feed $A_t$ back to Hawkes process as endogenous excitation $\phi_{int}$
 \ENDFOR
 \STATE Perform emotion contagion among social neighbors; update $B^{emo}$
 \STATE $\mathcal{T} \leftarrow \mathcal{T} \cup \{(t,\, \mathcal{A}_t,\, \{A_t\})\}$
\ENDFOR
\RETURN $\mathcal{T}$
\end{algorithmic}
\end{algorithm}

\subsection{Strategy Evaluation}\label{subsec:strategy}

Beyond simulation, POSIM also serves as a computational experimentation platform for strategy evaluation. Decision-makers often need to answer counterfactual questions such as ``how would opinion trajectories change under a given intervention?'' As shown in Figure~\ref{fig:framework}, strategy evaluation is organized around three modules (\textbf{Intervenor}, \textbf{Simulator}, and \textbf{Evaluator}), with the Simulator built on the agent--environment interaction loop described above.

\textbf{The Intervenor} connects external governance intent with the simulation state through three injection granularities. \textbf{Event Queue} broadcasts events to all agents at specified times (e.g., official statements). \textbf{Node Control} targets specific agents (e.g., directing an opinion leader to respond to a post). \textbf{Platform Policy} modifies environment parameters (e.g., adjusting recommendation weights or restricting content reach). These can be combined to provide intervention capability from individual to platform level.

\textbf{The Simulator} executes the interaction loop: agents perceive, decide, and act within the virtual platform, with outputs feeding back to the temporal engine. Injected events or policies propagate through this loop to affect subsequent cognition and behavior. A key mechanism is \textbf{checkpoint callback}: the system periodically saves state snapshots, allowing researchers to load any checkpoint, inject different interventions, and generate parallel trajectories from identical initial conditions, directly supporting counterfactual comparison.

\textbf{The Evaluator} is decoupled from the Simulator, reading simulation logs through standardized interfaces to perform quantitative assessment across hotness curves, sentiment distributions, opinion distributions, and topic migration. Results directly inform subsequent intervention design.

\textbf{LLM resource scheduling.} Large-scale simulation requires massive LLM calls. The framework manages multiple endpoints through a unified API pool that maintains per-usage-type model counters (belief update, desire generation, action decision) to balance reasoning quality and cost. To prevent homogenized outputs, the system perturbs sampling parameters: $T' = T + \epsilon_T$ ($\epsilon_T \sim \mathcal{U}(-0.1, 0.1)$), $p' = p + \epsilon_p$ ($\epsilon_p \sim \mathcal{U}(-0.05, 0.05)$), with random frequency and presence penalties for diversity. The pool also provides concurrency control, round-robin load balancing, fault detection, and rate limiting.

\textbf{Modular architecture.} Different research scenarios demand different cognitive models, temporal dynamics, and evaluation dimensions. POSIM therefore follows the \textbf{separation of concerns} principle, encapsulating agent cognition, environment, and strategy evaluation as independent modules that communicate through structured states rather than shared implementation. Researchers can replace the cognitive architecture, switch the temporal engine, or add evaluation dimensions without affecting other modules. This modular design positions POSIM as an extensible platform for diverse social simulation research.

\section{Validation Framework}\label{sec:validation}

Our validation framework builds on the classical V\&V (Verification and Validation) principles from systems engineering~\citep{sargent_verification_2013}, which distinguish \textit{verification} (``Have we built the model correctly?'') from \textit{validation} (``Are the model's results accurate?''). Following this distinction, we design a three-tier progressive evaluation. The first two tiers, individual behavioral mechanism calibration and collective phenomenon emergence calibration, correspond to \textit{verification}, examining whether the simulation's internal mechanisms work as intended and produce plausible emergent behaviors. The third tier, statistical result consistency calibration, corresponds to \textit{validation}, examining whether simulation outputs statistically match real data.

\textbf{Individual behavioral mechanism calibration} (Section~\ref{subsec:micro}) verifies whether agents make cognitively sound decisions as designed, serving as a decision quality check. This tier examines mechanism correctness along three dimensions. Cognitive-behavior chain consistency checks whether the pipeline from belief update to behavioral output is complete, coherent, and logically grounded. Personality stability assesses whether agents maintain role-consistent behavioral styles across interaction rounds. Decision robustness tests whether decisions remain stable under semantically equivalent input perturbations. Individual-level mechanism correctness is the foundation of the entire validation hierarchy: only when individual decision mechanisms are sound can collective emergence be meaningfully interpreted.

\textbf{Collective phenomenon emergence calibration} (Section~\ref{subsec:macro}) verifies whether sound individual mechanisms give rise to emergent group phenomena consistent with social science theory. This tier treats the simulation as a computational experiment, examining whether the system spontaneously produces theoretically predicted collective patterns without explicit group-level programming, including multi-phase public opinion lifecycle evolution, systematic behavioral heterogeneity across agent types, emotional arousal and polarization, and scale-free network topology with power-law cascade distributions. The appearance of these phenomena provides indirect but strong evidence for the soundness of individual-level mechanisms.

\textbf{Statistical result consistency calibration} (Section~\ref{subsec:stat}) further examines, after mechanism calibration, whether simulation outputs accurately reproduce real public opinion dynamics in a statistical sense. This tier compares simulation results with real data across three dimensions: behavior layer (action type distribution and behavioral hotness curves), content layer (discourse irrationality, lexical diversity, and sentiment tendency), and topology layer (network structure and information cascades), using nine quantitative metrics to establish statistical credibility.

\section{Experiments}\label{sec:exp}

\subsection{Datasets and Experimental Setup}

We collect three real-world public opinion event datasets from Sina Weibo, one of China's largest social media platforms. Its open interaction mechanisms (posting, reposting, commenting, and trending topics) make it a primary arena for public opinion formation and diffusion. Each dataset covers the full interaction chain of original posts, reposts, and comments. The three events span social controversy, campus incident, and food safety categories, covering diverse opinion patterns from short-term outbreaks to long-cycle multi-phase evolution. The \textbf{Luxury Earring event} (LE) occurred in May 2025, when jewelry worn by a public figure in a widely circulated photo was identified as a luxury item, triggering widespread public discussion. The topic repeatedly trended on Weibo, showing a typical short-term concentrated outbreak pattern. The \textbf{WHU Library event} (WL) originated from a reported harassment incident at Wuhan University's library in July 2023. The first-instance court verdict in July 2025 reignited extensive public discourse, resulting in a long-cycle multi-phase evolution from dormancy to resurgence. The \textbf{Xibei Prepared Food event} (XF) was triggered in September 2025 when an internet personality alleged that a restaurant chain used prepared food in its dishes. The ensuing debate over food safety and consumer rights generated sustained public attention, showing a multi-party contention pattern with prolonged fermentation.

We select the peak period of each event for simulation, maximizing information density under limited computational resources. Simulation start times are: LE 2025-05-16 08:00, WL 2025-07-27 08:00, XF 2025-09-14 05:00; duration and step counts are listed in Table~\ref{tab:dataset}. Descriptive statistics after deduplication, low-activity user filtering, and irrelevant content removal are also shown in Table~\ref{tab:dataset}, with detailed collection and preprocessing procedures in Appendix~\ref{app:data}. The temporal resolution is 10 minutes per step. Agent beliefs are initialized from each user's historical posts before the simulation start time, sequentially constructing role identity, psychological cognition, event opinion, and initial emotion vector to form a complete personalized Social-BDI belief system.

\begin{table}[!t]
\centering
\caption{Descriptive statistics of three public opinion event datasets.}
\label{tab:dataset}
\footnotesize
\renewcommand{\arraystretch}{1.05}
\begin{tabular}{@{}lccccc@{}}
\toprule
\textbf{Event} & \textbf{\#Users} & \textbf{\#Posts} & \textbf{Duration} & \textbf{\#Steps} & \textbf{Category} \\
\midrule
LE & 1,530 & 34,218 & ${\sim}$46h & 276 & Social Controversy \\
WL & 1,843 & 51,647 & ${\sim}$190h & 1,140 & Campus Incident \\
XF & 1,987 & 14,892 & ${\sim}$71h & 426 & Food Safety \\
\bottomrule
\end{tabular}
\par\vspace{2pt}
{\scriptsize\rmfamily
\textit{Note}: LE = Luxury Earring; WL = WHU Library; XF = Xibei Prepared Food.}
\end{table}

\subsection{Individual Behavioral Mechanism Calibration}\label{subsec:micro}

\subsubsection{Baselines}

Individual mechanism validation focuses on single-agent decision quality. Baselines are compared under identical input conditions to examine how different cognitive architectures affect decision interpretability, personality consistency, and robustness. Four methods are evaluated. \textbf{Direct-Nothink} uses the non-thinking model Qwen2.5-7B-Instruct\footnote{\url{https://huggingface.co/Qwen/Qwen2.5-7B-Instruct}}, concatenating environment information and role description into a single prompt for direct behavior output without explicit cognitive reasoning. \textbf{Direct-Think} upgrades to the stronger thinking model Qwen3-8B\footnote{\url{https://huggingface.co/Qwen/Qwen3-8B}}, but still uses single-prompt direct output. \textbf{CoT} uses the same base model as Social-BDI (Qwen2.5-7B-Instruct), completing belief inference, desire analysis, and behavior decision within a single LLM call via chain-of-thought prompting. The three stages are serialized within one context window rather than modularized independently. \textbf{Social-BDI} is the proposed method, also using Qwen2.5-7B-Instruct, which decomposes the cognitive process into three independent modules (belief update, desire generation, intention planning), each completed by a separate LLM call with structured intermediate state transfer.

\subsubsection{Evaluation Setup and Results}

We randomly sample $N=500$ users as agents and run $T=12$ rounds, with all four methods operating under identical conditions. Three evaluation dimensions are assessed. \textbf{Cognitive-behavior chain consistency} is rated on a 0 to 5 scale across four sub-dimensions: chain completeness, emotional dynamics rationality, motivation-behavior alignment, and information integration. \textbf{Personality stability} evaluates the similarity (0 to 1) between the agent's ground-truth identity description and that reverse-inferred by an LLM from behavioral trajectories alone. \textbf{Decision robustness} measures decision stability (0 to 1) under semantically equivalent input perturbations. All three evaluations use a fixed-hyperparameter LLM judge (prompt templates in Appendix~\ref{app:eval_prompts}). Results are shown in Table~\ref{tab:micro}.

\begin{table}[!t]
\centering
\caption{Micro-level behavioral mechanism validation results (mean$\pm$std).}
\label{tab:micro}
\renewcommand{\arraystretch}{1.05}
\resizebox{\columnwidth}{!}{%
\scriptsize
\begin{tabular}{@{}lccc@{}}
\toprule
\textbf{Method} & \makecell{\textbf{Cognitive-Behavior}\\\textbf{Chain Consistency} (0--5)\,$\uparrow$} & \makecell{\textbf{Personality}\\\textbf{Stability} (0--1)\,$\uparrow$} & \makecell{\textbf{Decision}\\\textbf{Robustness} (0--1)\,$\uparrow$} \\
\midrule
Direct-Nothink & 1.47$\pm$0.50 & 0.478$\pm$0.263 & 0.629$\pm$0.240 \\
Direct-Think & 1.75$\pm$0.43 & 0.448$\pm$0.269 & 0.603$\pm$0.299 \\
CoT & 3.09$\pm$0.29 & 0.516$\pm$0.272 & 0.541$\pm$0.356 \\
\rowcolor{grayrow}
\textbf{Social-BDI (Ours)} & \textbf{4.64}$\pm$\textbf{0.48} & \textbf{0.661}$\pm$\textbf{0.215} & \textbf{0.695}$\pm$\textbf{0.213} \\
\bottomrule
\end{tabular}%
}
\par\vspace{2pt}
{\scriptsize\rmfamily
\textit{Note}: Cognitive-Behavior Chain Consistency is rated on a 0--5 scale; Personality Stability and Decision Robustness are scored on a 0--1 scale. \textbf{Bold}: best per metric.}
\end{table}

Social-BDI outperforms all baselines across all three dimensions. Its chain consistency score of 4.64 significantly exceeds CoT's 3.09 ($p < 0.001$), confirming that the modular architecture produces more complete and logically grounded decision chains. Personality stability and decision robustness are likewise highest, validating the belief grounding mechanism. Notably, despite using a stronger thinking model, Direct-Think performs worse than Direct-Nothink on personality stability and robustness, showing that raw reasoning capability alone cannot substitute for explicit cognitive architecture. CoT serializes cognitive reasoning in a single call but yields the lowest robustness, suggesting that single-call serialization lacks stable state anchoring, whereas Social-BDI's explicit belief states provide effective cognitive grounding.

\subsection{Collective Phenomenon Emergence Calibration}\label{subsec:macro}

\subsubsection{Public Opinion Lifecycle}

\begin{figure}[!t]
\centering
\includegraphics[width=\columnwidth]{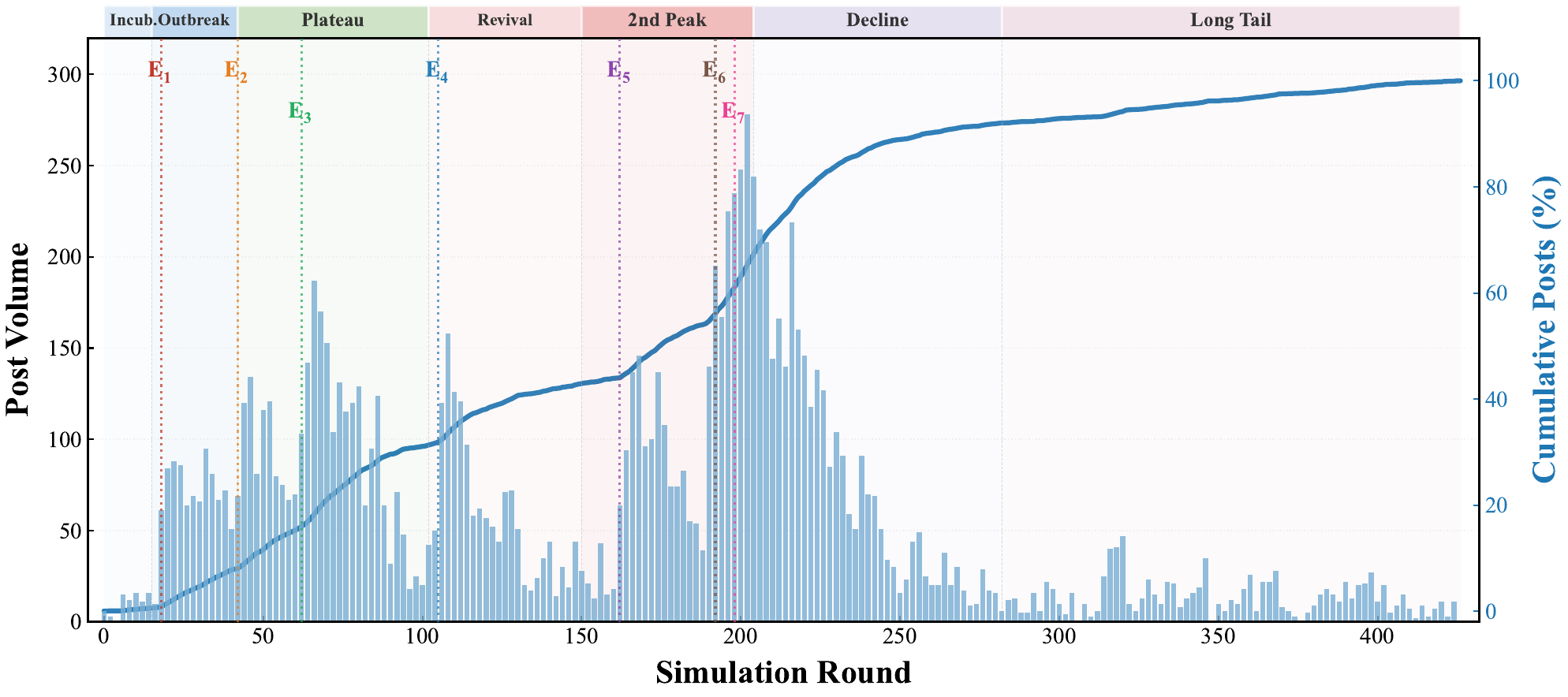}
\caption{Simulated public opinion lifecycle, showing posting volume (bars, left axis) and cumulative posting percentage S-curve (solid line, right axis). $E_1$--$E_7$ mark exogenous event injection points.}
\label{fig:lifecycle}
\end{figure}

Public opinion lifecycle theory holds that online public opinion events do not evolve at a uniform pace but exhibit distinct phase-dependent patterns~\citep{guo_stochastic_2024,chen_evolution_2025}. A typical lifecycle includes an outbreak phase (concentrated public attention with rapidly rising activity), a plateau phase (stabilized or mildly fluctuating hotness), possible resurgence (secondary outbreaks from new stimuli), and a decay phase (attention shift and activity decline). Not all events follow the same phase sequence: short-burst events may proceed directly from outbreak to decay, while multi-party contention events may undergo multiple resurgence cycles.

Taking the XF event as an illustrative example (all emergence phenomena use this event), Figure~\ref{fig:lifecycle} shows that the simulation clearly exhibits a multi-phase lifecycle from outbreak through plateau and resurgence to decay, with each phase transition traceable to specific exogenous events. After $E_1$ (kitchen video exposure), posting volume surges rapidly, entering the outbreak phase. $E_2$ (temporary ceasefire announcement) leads to a noticeable activity decline, entering the plateau phase. $E_3$ (industry debate on prepared food definitions) sustains moderate discussion without triggering a large-scale outbreak. $E_4$ (founder's leaked group chat screenshot) reignites discussion, breaking the ceasefire. $E_5$ (media collective commentary) triggers the peak activity of the entire simulation, showing a classic resurgence pattern where official media framing triggers a secondary outbreak. The cumulative posting percentage follows an S-curve, consistent with diffusion theory in communication studies~\citep{kong_exploring_2020}.

Crucially, this multi-phase lifecycle is not driven by preset rules but emerges spontaneously from two cooperating mechanisms. At the \textbf{environment level}, the Hawkes process self-exciting mechanism causes exogenous event injection to trigger an ``excitation, amplification, decay'' dynamic response. Exogenous events elevate base activation intensity, and event-related content published by agents further amplifies collective activity through the self-exciting term $\sum \alpha e^{-\beta(t-t_i)}$, forming a positive feedback loop. When event stimulation fades and self-exciting effects decay, activity naturally subsides. At the \textbf{agent level}, the Social-BDI architecture produces differentiated cognitive responses to different event types. Food safety crises ($E_1$) and personal attack screenshots ($E_4$) trigger strong emotional arousal and belief conflict, driving high-density participation, whereas technical industry debates ($E_3$) produce weaker cognitive impact on ordinary users, sustaining only moderate discussion. This synergy between environment-driven and cognition-driven dynamics enables the simulation to naturally reproduce the nonlinear mapping between event types and public response intensity observed in real public opinion events.

\subsubsection{Multi-Agent Behavioral Heterogeneity}

\FloatBarrier
\begin{figure}[!t]
\centering
\includegraphics[width=\columnwidth]{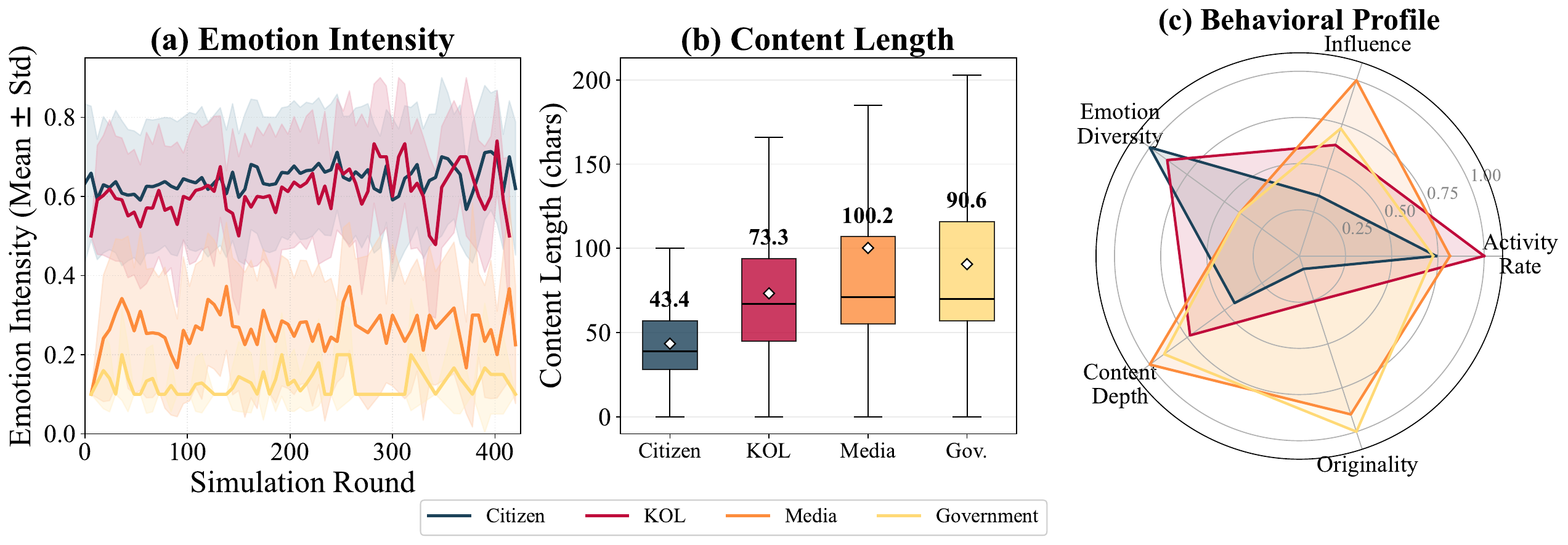}
\caption{Behavioral heterogeneity across four agent types. (a)~Emotional intensity temporal evolution; (b)~Content length distribution; (c)~Multi-dimensional behavioral profile radar chart.}
\label{fig:heterogeneity}
\end{figure}

Figure~\ref{fig:heterogeneity} illustrates the systematic heterogeneity across four agent types in emotional intensity, content length, and multi-dimensional behavioral profiles. For emotional intensity, ordinary user and opinion leader agents maintain average intensities of 0.645 and 0.603 respectively, consistently in the high-arousal range with sharper post-event fluctuations. Media accounts and governments remain in the low-arousal range, showing a ``public emotionality, official neutrality'' stratification consistent with real opinion observations. For content length, media accounts and governments produce significantly longer outputs than ordinary users, reflecting more structured content generation. The radar chart reveals four distinct profiles: ordinary users are prominent in emotion and interaction frequency, opinion leaders lead in influence and content volume, media accounts excel in content quality and neutral expression, and governments are characterized by low-frequency but high-authority contributions. These heterogeneity patterns are not manually coded but emerge naturally from differentiated role beliefs and motivation weights within the Social-BDI architecture, validating its effectiveness in modeling multi-agent heterogeneity.

\subsubsection{Emotional Arousal and Polarization}

Emotional arousal theory posits that high-arousal states (e.g., anger, excitement) more strongly motivate social sharing and information propagation than low-arousal states (e.g., sadness, contentment)~\citep{berger_what_2012}. Emotional contagion theory holds that individuals automatically mimic and synchronize others' emotional expressions during social interaction, causing emotions to spread through groups~\citep{hatfield_emotional_1994}. These two dimensions describe the \textbf{intra-individual reinforcement} and \textbf{inter-individual diffusion} mechanisms of emotion, jointly forming the micro-foundation of opinion sentiment evolution. We therefore validate along both dimensions (Table~\ref{tab:emotion}). The arousal self-reinforcement dimension includes three metrics: (1)~High-arousal trend coefficient~$\beta$, obtained via linear regression on high-arousal emotion proportion, where $\beta > 0$ indicates an upward trend; (2)~High-arousal emotion ratio, the overall proportion of high-arousal emotions; (3)~High/low arousal intensity ratio. The emotional contagion dimension includes: (4)~Comment-chain emotion consistency, the proportion of adjacent comments sharing the same emotional polarity; (5)~Escalation/de-escalation ratio, the ratio of low-to-high versus high-to-low arousal transitions, where values above~1 indicate a ratchet-like accumulation pattern.

\begin{table}[!t]
\centering
\caption{Emotional arousal theory validation metrics.}
\label{tab:emotion}
\renewcommand{\arraystretch}{1.0}
\resizebox{\columnwidth}{!}{%
\scriptsize
\begin{tabular}{@{}p{1.5cm}p{3.0cm}cc@{}}
\toprule
\textbf{Dimension} & \textbf{Metric} & \textbf{Sim.\ Result} & \textbf{Theoretical Expectation} \\
\midrule
\multirow{3}{1.5cm}{Arousal Self-Reinforcement$^{\rm a}$}
 & High-arousal trend coeff.\ $\beta$ & $3.41{\times}10^{-4}$\,($p{<}.001$) & $\beta > 0$ \\
 & High-arousal emotion ratio & 73.5\% & Dominant \\
 & High/low arousal intensity ratio & 2.0$\times$ & $>1$ \\
\midrule
\multirow{2}{1.5cm}{Emotional Contagion$^{\rm b}$}
 & Comment-chain emotion consistency & 0.772 & $\gg$ random baseline \\
 & Escalation/de-escalation ratio & 4.78 & $>1$ (ratchet effect) \\
\bottomrule
\end{tabular}%
}
\vspace{3pt}
{\footnotesize $^{\rm a}$~\citet{berger_what_2012};\quad $^{\rm b}$~\citet{hatfield_emotional_1994}}
\end{table}

\begin{figure}[!t]
\centering
\includegraphics[width=0.85\columnwidth]{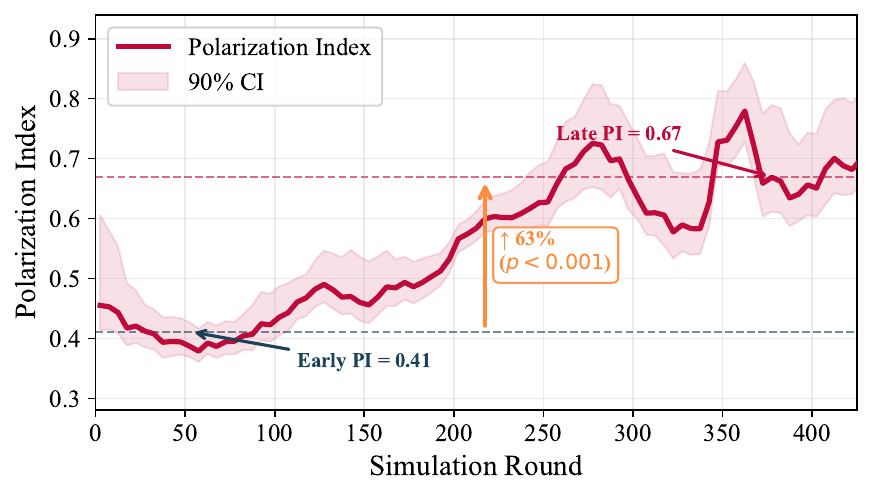}
\caption{Emotion polarization index (PI) evolution over simulation rounds. Solid line: PI mean; shading: 90\% bootstrap confidence interval; dashed lines: early and late stage means.}
\label{fig:polarization}
\end{figure}

Results show that the high-arousal emotion proportion rises significantly over time, reaching 73.5\%, validating the arousal self-reinforcement effect. Comment-chain emotion consistency reaches 0.772, and the escalation-to-de-escalation ratio is 4.78, indicating that emotions in interactions tend to intensify rather than subside. This ratchet-like accumulation pattern is consistent with the emotional contagion hypothesis~\citep{hatfield_emotional_1994}.
We further define a polarization index:
\begin{equation}
\text{PI} = 1 - \frac{H}{H_{\max}}, \quad H = -\sum_{k} p_k \ln p_k
\end{equation}
where $H$ is Shannon entropy over the emotion distribution, $p_k$ is the proportion of emotion~$k$, and $H_{\max}$ is maximum entropy under uniform distribution. As shown in Figure~\ref{fig:polarization}, PI rises from 0.41 in early stages to 0.67 in late stages (63\% increase, $p < 0.001$), indicating that the emotion distribution increasingly concentrates on a few dominant categories, consistent with theoretical expectations of emotional convergence in echo chamber environments~\citep{cinelli_echo_2021}.

\subsubsection{Scale-Free Topology and Cascade Power Law}

\begin{figure}[!t]
\centering
\includegraphics[width=\columnwidth]{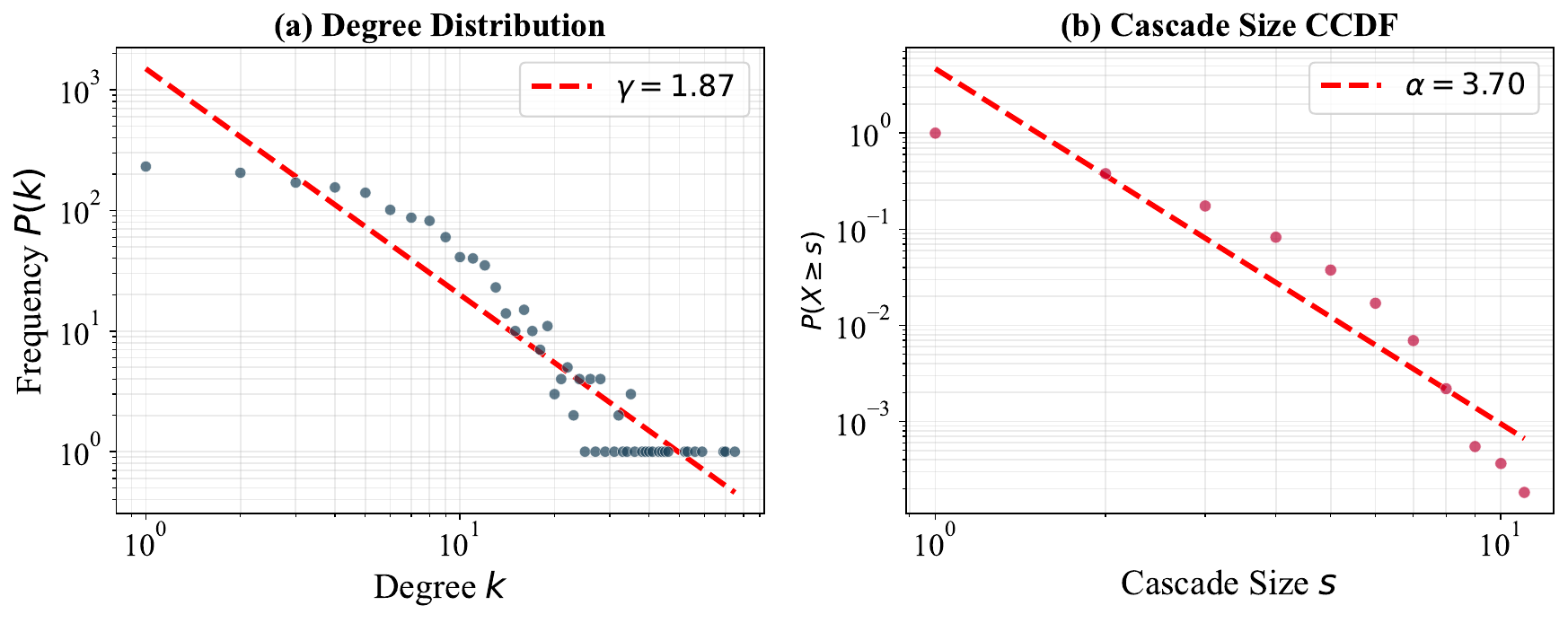}
\caption{Topological properties of the interaction network. (a)~Degree distribution power-law fit $P(k) \sim k^{-\gamma}$; (b)~Cascade size CCDF $P(X \geq s) \sim s^{-\alpha}$.}
\label{fig:powerlaw}
\end{figure}

Real social network degree distributions and information cascade sizes universally follow power-law distributions~\citep{barabasi_albert_1999}. As shown in Figure~\ref{fig:powerlaw}(a), the interaction network's degree distribution yields a power-law exponent $\gamma = 1.87$, within the $1.5 < \gamma < 3$ range commonly reported for real social networks. This indicates a typical scale-free topology where a few opinion leader nodes form high-degree hubs while most ordinary users have only sparse connections. Figure~\ref{fig:powerlaw}(b) shows the complementary cumulative distribution function (CCDF) of cascade sizes, $P(X \geq s) = 1 - F(s)$, which exhibits linearity on a log-log scale, indicating power-law behavior. The fitted exponent $\alpha = 3.70$ confirms that the vast majority of repost cascades are small, with occasional viral bursts involving large numbers of users, consistent with the ``most posts go unnoticed, a few go viral'' long-tail phenomenon observed on real Weibo.

\subsection{Statistical Result Calibration}\label{subsec:stat}

The core objective of public opinion simulation is to credibly reproduce real dynamics, which requires consistency across multiple dimensions. Behavior layer calibration verifies that agent participation patterns match real user activity, serving as the foundational premise of credibility. Content layer calibration examines whether generated text reproduces the discourse characteristics of real events (e.g., irrational expression, sentiment tendency), directly reflecting the simulation's ability to capture the semantic space. Topology layer calibration checks whether emergent interaction networks and cascades exhibit realistic structural properties, reflecting the reproduction of macroscopic information diffusion patterns. The three layers correspond to ``who participates,'' ``what is said,'' and ``how it propagates,'' forming a complete behavioral-to-structural hierarchy.

This section compares simulation outputs with three real events across \textbf{behavior layer} (action type distribution and behavioral hotness curves), \textbf{content layer} (discourse irrationality, lexical diversity, and sentiment tendency), and \textbf{topology layer} (network structure and information cascades).

\subsubsection{Baselines}

Statistical calibration examines whether large-scale multi-agent interactions produce emergent group behaviors that statistically align with real data, requiring end-to-end multi-step simulation in the complete environment. Three baselines are used. \textbf{Rule-based ABM} uses traditional rule-driven simulation where each agent samples behavior types from prior probability distributions based on historical behavior, with no LLM involvement; because it generates no natural-language text, content layer metrics are not applicable. \textbf{POSIM w/ Direct LLM} removes the Social-BDI architecture, corresponding to the Direct-Nothink setting in Section~\ref{subsec:micro} but embedded in the full simulation environment. Agents directly generate behavior and content in a single LLM step after receiving recommendations, without explicit belief update or desire reasoning. \textbf{POSIM w/ CoT} replaces Social-BDI's three independent modules with a single chain-of-thought call that serializes cognition to behavior in one LLM invocation, corresponding to the CoT setting in Section~\ref{subsec:micro}, retaining all other POSIM settings.

\subsubsection{Evaluation Metrics}

Evaluation compares simulation outputs with real data across behavior, content, and topology layers using nine metrics.

\textbf{Behavior layer} measures the consistency of agent group behavior patterns with real users, comprising three metrics. BType JSD uses Jensen--Shannon divergence to compare simulated and real action type distributions:
\begin{equation}
\text{JSD}(P \| Q) = \tfrac{1}{2} D_{\text{KL}}(P \| M) + \tfrac{1}{2} D_{\text{KL}}(Q \| M)
\end{equation}
where $M = (P + Q)/2$. Act.\,$\rho$ measures the Pearson correlation between simulated and real behavioral hotness curves. Act.\,RMSE measures their pointwise root mean squared error.

\textbf{Content layer} measures the semantic match between generated and real text, comprising three metrics. Irrat.\,Sim.\ classifies discourse into irrational, rational, and neutral categories via keyword matching and computes the similarity between simulated and real distributions. This directly reflects whether the simulation reproduces the emotion-driven irrational discourse landscape that the Social-BDI belief module is designed to capture. $|\Delta\text{TTR}|$ measures the difference in lexical richness (Type-Token Ratio) between simulated and real text. $|\Delta\bar{S}|$ measures the overall sentiment mean deviation, where $S = 2s - 1$ ($s \in [0,1]$ from SnowNLP\footnote{\url{https://github.com/isnowfy/snownlp}}) maps to $[-1, +1]$.

\textbf{Topology layer} measures the emergent network structure and propagation patterns, comprising three metrics. Net.\,Sim.\ compares simulated and real interaction network topological features. Casc.\,Sim.\ compares cascade size statistical distributions. Casc.\,PL compares cascade power-law exponents.

Layer averages are computed by normalizing lower-is-better metrics via $1{-}x$ and averaging with higher-is-better metrics.

\subsubsection{Calibration Results}

\begin{table*}[!t]
\centering
\caption{Statistical calibration results across three events (LE\,=\,Luxury Earring, WL\,=\,WHU Library, XF\,=\,Xibei Prepared Food). \textbf{Bold}: best per dataset; \na: not applicable. \textcolor{gaingreen}{Green}: gain of POSIM over the second-best method; gains are clipped to $+.000$ on metrics where POSIM does not rank first.}
\label{tab:calibration}
\vspace{2pt}
\renewcommand{\arraystretch}{1.05}
\resizebox{\textwidth}{!}{%
\footnotesize
\begin{tabular}{@{}cl cccc cccc cccc@{}}
\toprule
\multirow{2}{*}{\textbf{Data}} & \multirow{2}{*}{\textbf{Method}}
& \multicolumn{4}{c}{\textbf{Behavior Layer}}
& \multicolumn{4}{c}{\textbf{Content Layer}}
& \multicolumn{4}{c}{\textbf{Topology Layer}} \\
\cmidrule(lr){3-6} \cmidrule(lr){7-10} \cmidrule(lr){11-14}
& & \makecell{BType\\JSD\,$\downarrow$} & \makecell{Act.\\$\rho$\,$\uparrow$} & \makecell{Act.\\RMSE\,$\downarrow$} & \makecell{Avg.\\$\uparrow$}
& \makecell{Irrat.\\Sim.\,$\uparrow$} & \makecell{$|\Delta\text{TTR}|$\\$\downarrow$} & \makecell{$|\Delta\bar{S}|$\\$\downarrow$} & \makecell{Avg.\\$\uparrow$}
& \makecell{Net.\\Sim.\,$\uparrow$} & \makecell{Casc.\\Sim.\,$\uparrow$} & \makecell{Casc.\\PL\,$\uparrow$} & \makecell{Avg.\\$\uparrow$} \\
\midrule
\multirow{5}{*}{\textbf{LE}}
& Rule-based ABM        & 0.427 & 0.808 & 0.158 & 0.741 & \na & \na & \na & \na & 0.479 & 0.633 & 0.543 & 0.552 \\
& POSIM w/ Direct LLM        & 0.289 & 0.799 & 0.162 & 0.783 & 0.544 & 0.185 & 0.319 & 0.680 & 0.565 & 0.735 & 0.918 & 0.739 \\
& POSIM w/ CoT          & 0.394 & 0.806 & \best{0.151} & 0.754 & 0.584 & 0.141 & 0.123 & 0.774 & 0.755 & 0.777 & 0.756 & 0.763 \\
\rowcolor{oursrow}
& \textbf{POSIM (Ours)}  & \best{0.193} & \best{0.809} & 0.154 & \best{0.821} & \best{0.790} & \best{0.030} & \best{0.029} & \best{0.910} & \best{0.895} & \best{0.830} & \best{0.961} & \best{0.896} \\[1pt]
\rowcolor{oursrow}
& \textit{Improvement} & \gain{+.096} & \gain{+.001} & \gain{+.000} & \gain{+.038} & \gain{+.206} & \gain{+.111} & \gain{+.094} & \gain{+.136} & \gain{+.140} & \gain{+.053} & \gain{+.043} & \gain{+.133} \\
\midrule
\multirow{5}{*}{\textbf{WL}}
& Rule-based ABM        & 0.318 & 0.681 & 0.126 & 0.746 & \na & \na & \na & \na & 0.869 & 0.696 & 0.642 & 0.736 \\
& POSIM w/ Direct LLM        & 0.237 & 0.722 & 0.119 & 0.789 & 0.453 & 0.172 & 0.360 & 0.640 & 0.528 & 0.667 & 0.580 & 0.592 \\
& POSIM w/ CoT          & 0.229 & 0.744 & \best{0.116} & 0.800 & 0.474 & 0.091 & 0.365 & 0.673 & \best{0.941} & 0.740 & 0.673 & 0.784 \\
\rowcolor{oursrow}
& \textbf{POSIM (Ours)}  & \best{0.073} & \best{0.750} & 0.118 & \best{0.853} & \best{0.841} & \best{0.010} & \best{0.203} & \best{0.876} & 0.850 & \best{0.758} & \best{0.965} & \best{0.858} \\[1pt]
\rowcolor{oursrow}
& \textit{Improvement} & \gain{+.156} & \gain{+.006} & \gain{+.000} & \gain{+.053} & \gain{+.367} & \gain{+.081} & \gain{+.157} & \gain{+.203} & \gain{+.000} & \gain{+.018} & \gain{+.292} & \gain{+.074} \\
\midrule
\multirow{5}{*}{\textbf{XF}}
& Rule-based ABM        & 0.312 & 0.664 & 0.187 & 0.721 & \na & \na & \na & \na & 0.528 & 0.614 & 0.279 & 0.474 \\
& POSIM w/ Direct LLM        & 0.244 & 0.671 & 0.190 & 0.746 & 0.765 & 0.117 & 0.076 & 0.858 & 0.774 & 0.695 & 0.482 & 0.650 \\
& POSIM w/ CoT          & 0.293 & 0.699 & 0.181 & 0.742 & 0.774 & 0.134 & \best{0.014} & 0.875 & 0.767 & \best{0.696} & 0.460 & 0.641 \\
\rowcolor{oursrow}
& \textbf{POSIM (Ours)}  & \best{0.148} & \best{0.727} & \best{0.168} & \best{0.804} & \best{0.843} & \best{0.019} & 0.046 & \best{0.926} & \best{0.885} & \best{0.696} & \best{0.513} & \best{0.698} \\[1pt]
\rowcolor{oursrow}
& \textit{Improvement} & \gain{+.096} & \gain{+.028} & \gain{+.013} & \gain{+.058} & \gain{+.069} & \gain{+.098} & \gain{+.000} & \gain{+.051} & \gain{+.111} & \gain{+.000} & \gain{+.031} & \gain{+.048} \\
\bottomrule
\end{tabular}%
}
\end{table*}

\begin{figure*}[!t]
\centering
\includegraphics[width=0.75\textwidth]{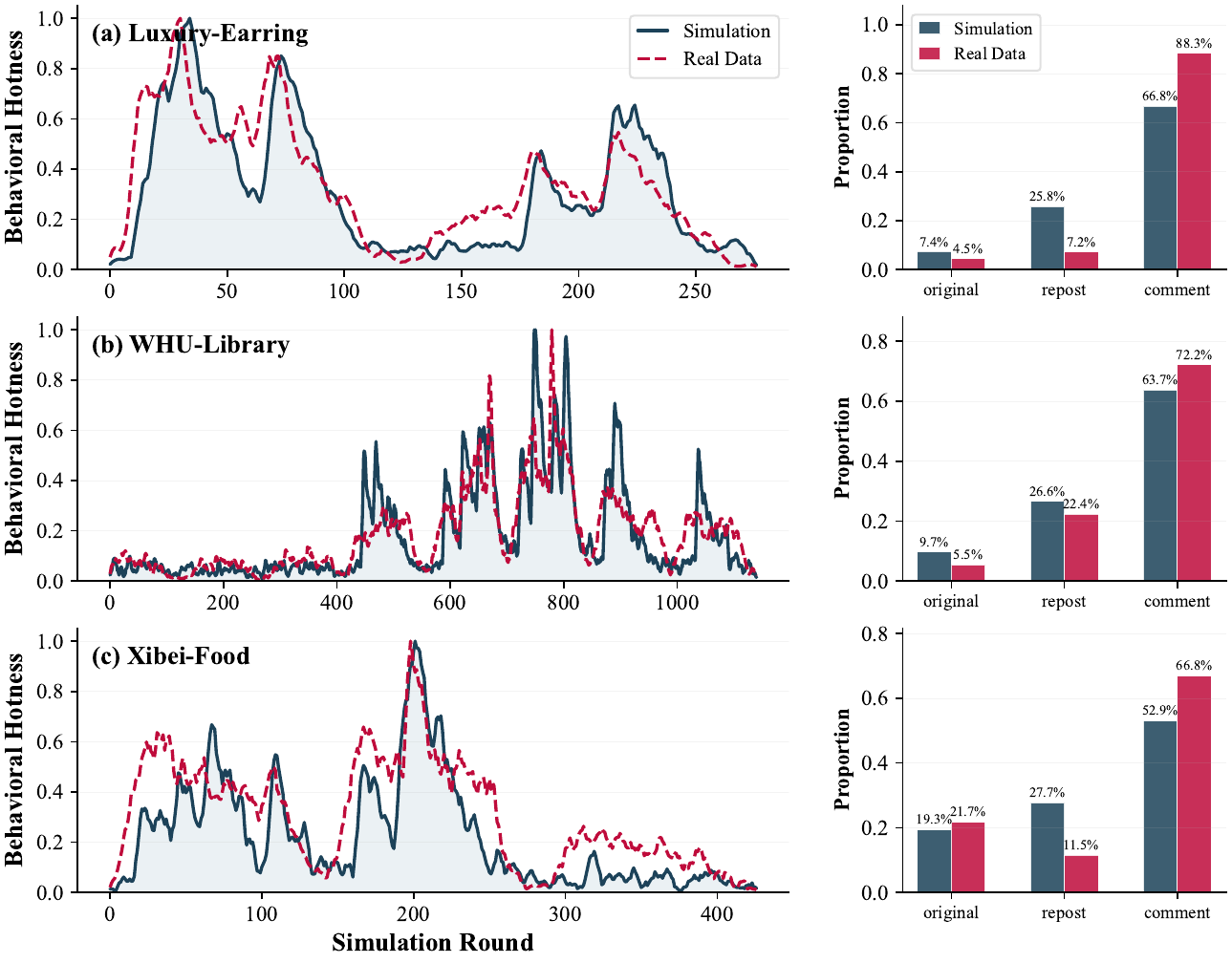}
\caption{Behavioral hotness and distribution calibration results for three events. Each row corresponds to one event ((a)~Luxury Earring, (b)~WHU Library, (c)~Xibei Prepared Food); left panels show simulated vs.\ real behavioral hotness curves, right panels show action type proportion comparisons.}
\label{fig:three_event_calibration}
\end{figure*}

Experiments are conducted on all three datasets, with results shown in Table~\ref{tab:calibration} and Figure~\ref{fig:three_event_calibration}.

\textbf{Behavior layer calibration.} Figure~\ref{fig:three_event_calibration} shows POSIM's behavioral hotness curves and action type distribution calibration across all three events. POSIM achieves the best BType JSD on all datasets, indicating that Social-BDI produces action type proportions highly consistent with real data. WL's JSD of only 0.073 is far superior to other methods, likely because the relatively concentrated behavioral patterns in campus discussions are well captured by Social-BDI's belief grounding. Single-step LLM decisions produce significantly larger distribution deviations, suggesting that LLMs without layered cognitive constraints tend toward action type homogenization. For activity correlation~$\rho$, POSIM maintains high values across all datasets, confirming that the Hawkes-driven temporal engine effectively captures temporal fluctuation patterns. POSIM's Act.\,RMSE is slightly behind CoT on some datasets, possibly reflecting a trade-off between pointwise precision and overall distributional consistency, but POSIM leads on composite metrics. Behavior layer averages reach 0.821, 0.853, and 0.804 across the three datasets, with stable cross-event performance demonstrating framework generality.

\textbf{Content layer calibration.} POSIM achieves the highest irrationality similarity on all datasets, indicating that Social-BDI accurately reproduces the rational-irrational discourse landscape of real public opinion events. This advantage stems from the belief subsystem's explicit modeling of emotional arousal and cognitive biases, which enables agents to generate irrational expressions consistent with their cognitive states. In contrast, POSIM w/ Direct LLM shows markedly lower irrationality similarity, especially on LE, because the absence of belief-driven cognition causes LLM outputs to tend toward neutral expression. Lexical diversity deviation $|\Delta\text{TTR}|$ is smallest on all datasets, indicating that personalized role information maintained by the belief module effectively mitigates the LLM's ``average personality'' problem, yielding individually styled text across agents. Sentiment deviation $|\Delta\bar{S}|$ remains minimal on most datasets. On XF, CoT's sentiment deviation is slightly lower than POSIM's, but its irrationality similarity and lexical diversity are both inferior, suggesting this local advantage likely reflects coincidental numerical proximity rather than more accurate sentiment modeling. Content layer averages are 0.910, 0.876, and 0.926.

\textbf{Topology layer calibration.} Using provenance chain tracking, POSIM achieves the best topology layer averages across all datasets. For network similarity, POSIM significantly outperforms Rule-based ABM on LE and XF, indicating that Social-BDI's multi-level intention planning enables agents to make selective interaction choices based on content quality and social affinity, producing more realistic network topology. On WL, CoT's network similarity slightly exceeds POSIM's, possibly because the campus-focused discussion yields a relatively simple network structure. However, CoT trails POSIM on cascade power-law and cascade size, indicating its network-level advantage does not extend to propagation patterns. POSIM's cascade power-law advantage is particularly pronounced, as Rule-based ABM's random interaction decisions cannot produce the long-tail cascade distributions of real social networks. XF's relatively lower topology averages likely reflect the event's longer duration and more complex network structure, but POSIM still significantly outperforms all baselines.

\textbf{Overall analysis.} POSIM leads comprehensively across all three layers. The core advantages are rooted in Social-BDI's layered cognitive design. The belief subsystem provides personalized cognitive grounding for decisions, the desire subsystem drives accurate reproduction of irrational discourse through motivational constraints, and the intention subsystem's structured decomposition ensures lexical diversity and interaction topology quality. Baseline performance differences corroborate this interpretation. Rule-based ABM lacks cognitive drive and performs poorly on topology in most datasets. POSIM w/ Direct LLM lacks belief grounding and shows insufficient content consistency. POSIM w/ CoT lacks inter-module independence and falls in between.

\vspace{6pt}
\subsection{Ablation Study}

To verify the necessity of each core module, we conduct ablation experiments on the LE dataset and compare the full framework with several variant configurations. \textbf{Full POSIM} includes all modules. \textbf{w/o Belief} removes the belief update module. Agents maintain no explicit belief state, with desire and intention generated directly from current perception without historical cognitive accumulation. \textbf{w/o Desire} removes desire generation. Intention planning receives no motivational weight constraints, and the LLM generates behavior directly from beliefs without intermediate motivational reasoning. \textbf{w/o Intention} removes intention planning. Behavior is no longer decomposed through multi-level chain-of-thought (action selection, expression strategy, content generation), and the LLM generates final output directly from beliefs and desires. \textbf{w/o Hawkes} replaces the Hawkes temporal engine with fixed-step uniform activation, activating a constant number of agents per step and removing event-driven dynamic scheduling. Results are shown in Table~\ref{tab:ablation}.

\begin{table}[!t]
\centering
\caption{Ablation study results. \underline{Underline} marks cases where a variant outperforms the full model.}
\label{tab:ablation}
\renewcommand{\arraystretch}{1.0}
\resizebox{\columnwidth}{!}{%
\scriptsize
\begin{tabular}{@{}l ccc ccc cc@{}}
\toprule
\multirow{2}{*}{\textbf{Configuration}} & \multicolumn{3}{c}{\textbf{Behavior Layer}} & \multicolumn{3}{c}{\textbf{Content Layer}} & \multicolumn{2}{c}{\textbf{Topology Layer}} \\
\cmidrule(lr){2-4} \cmidrule(lr){5-7} \cmidrule(lr){8-9}
& \makecell{JSD\\$\downarrow$} & \makecell{$\rho$\\$\uparrow$} & \makecell{RMSE\\$\downarrow$}
& \makecell{Irrat.\\$\uparrow$} & \makecell{$|\Delta$TTR$|$\\$\downarrow$} & \makecell{$|\Delta\bar{S}|$\\$\downarrow$}
& \makecell{Net.\\$\uparrow$} & \makecell{Casc.\\$\uparrow$} \\
\midrule
\rowcolor{grayrow}
Full POSIM & \best{0.193} & \best{0.809} & \best{0.154} & \best{0.790} & \best{0.030} & \best{0.029} & \best{0.895} & \best{0.830} \\
w/o Belief & 0.258 & 0.762 & 0.172 & 0.706 & 0.058 & 0.067 & 0.861 & 0.773 \\
w/o Desire & 0.267 & 0.779 & 0.169 & 0.682 & 0.071 & 0.083 & 0.853 & 0.788 \\
w/o Intention & 0.237 & 0.802 & 0.159 & 0.728 & 0.064 & 0.055 & 0.858 & 0.814 \\
w/o Hawkes & \underline{0.177} & 0.235 & 0.362 & 0.787 & 0.207 & \underline{0.028} & 0.822 & 0.754 \\
\bottomrule
\end{tabular}%
}
\par\vspace{2pt}
{\scriptsize\rmfamily
\textit{Note}: \textbf{Bold}: best per metric; \underline{underline}: variant outperforms the full model. Casc.\ combines cascade size and power-law similarity.}
\end{table}

\textbf{w/o Hawkes.} Removing the Hawkes process causes $\rho$ to plummet from 0.809 to 0.235 and RMSE to deteriorate from 0.154 to 0.362, confirming that uniform activation cannot capture temporal fluctuations. Its BType JSD (0.177) is marginally better than Full POSIM (0.193) because uniform activation keeps per-step agent counts constant, causing action type proportions to converge to their intrinsic mean under heavy sampling. This is a trivial advantage obtained at the cost of losing event-driven temporal dynamics. $|\Delta\text{TTR}|$ worsens to 0.207, indicating that temporal engine disruption indirectly causes content homogenization. $|\Delta\bar{S}|$ (0.028) is marginally better than Full POSIM (0.029). Uniform activation flattens event-driven sentiment peaks and troughs, causing the overall mean to coincidentally approach the real data's aggregate mean, but at the expense of entirely losing temporal sentiment dynamics.

\textbf{w/o Belief.} Removing beliefs causes broad degradation. JSD worsens to 0.258, irrationality similarity drops from 0.790 to 0.706, and $|\Delta\bar{S}|$ rises to 0.067. The belief module's $B^{evt}$ and $B^{emo}$ are the core drivers of irrational discourse generation. Their absence strips agents of deep cognition about contentious focal points and emotional arousal.

\textbf{w/o Desire.} Removing desires causes the most severe content layer degradation. Irrationality similarity drops to 0.682 (lowest among all variants) and $|\Delta\bar{S}|$ worsens to 0.083 (highest), indicating that motivation types are the core driver of irrational discourse expression. Without them, the LLM defaults to neutral text.

\textbf{w/o Intention.} Removing intention planning causes $|\Delta\text{TTR}|$ to worsen to 0.064 and network similarity to drop by 3.7pp, indicating that multi-level chain-of-thought decomposition substantially contributes to lexical diversity and interaction topology quality.

\textbf{Overall analysis.} The four ablation experiments demonstrate clear functional specialization. The belief subsystem provides global cognitive support, the desire subsystem drives content-layer irrational expression, the intention subsystem ensures lexical diversity and topology quality, and the Hawkes engine governs temporal consistency. Together, they validate the complementarity of Social-BDI's layered architecture.

\section{Discussion}\label{sec:discussion}

This section demonstrates the application value of POSIM as a computational experimentation platform through two case studies, then summarizes key findings and discusses limitations.

\subsection{Cognitive Priming Experiment}

A central challenge in public opinion governance is how to effectively guide public cognition to mitigate negative sentiment diffusion. However, directly testing cognitive intervention strategies in real environments faces dual constraints of ethics and controllability. Simulation platforms offer a risk-free computational experimentation pathway. This experiment evaluates the impact of different cognitive priming strategies on collective emotional evolution, providing quantitative evidence for cognitive intervention in public opinion governance.

We design a cognitive priming experiment based on the Luxury Earring event. Starting from a time step after an event burst during the simulation, we modify the psychological cognition beliefs ($B^{psy}$) of a proportion (coverage) of agents to simulate and observe how populations with different cognitive orientations respond to the event, testing three coverage levels: 30\%, 60\%, and 100\%. The control group maintains the no-intervention baseline. Three intervention strategies correspond to different cognitive regulation mechanisms. \textbf{Rational Cognition} (RC) guides agents toward multi-perspective rational analysis. \textbf{Empathy Priming} (EP) encourages understanding the parties' positions. \textbf{Emotional Regulation} (ER) guides agents to recognize and manage their own emotions. Results are shown in Table~\ref{tab:priming}.

\begin{table}[!t]
\centering
\caption{Cognitive priming experiment results. Coverage denotes the proportion of agents receiving cognitive priming.}
\label{tab:priming}
\scriptsize
\renewcommand{\arraystretch}{1.0}
\begin{tabular}{@{}llccc@{}}
\toprule
\textbf{Condition} & \textbf{Coverage} & \makecell{\textbf{Neg.\ Emotion}\\\textbf{Ratio}\,$\downarrow$} & \makecell{\textbf{Anger}\\\textbf{Ratio}\,$\downarrow$} & \makecell{\textbf{Emotion}\\\textbf{Intensity}\,$\downarrow$} \\
\midrule
Control & --- & 0.844 & 0.600 & 0.649 \\
\midrule
 & 30\% & 0.767 & 0.534 & 0.621 \\
 & 60\% & 0.753 & 0.506 & 0.595 \\
\rowcolor{grayrow}
\multirow{-3}{*}{\makecell[l]{Rational\\Cognition (RC)}} & 100\% & \best{0.571} & \best{0.407} & \best{0.527} \\
\midrule
\multirow{3}{*}{\makecell[l]{Empathy\\Priming (EP)}}
 & 30\% & 0.794 & 0.523 & 0.609 \\
 & 60\% & 0.806 & 0.535 & 0.617 \\
 & 100\% & 0.878 & 0.569 & 0.650 \\
\midrule
\multirow{3}{*}{\makecell[l]{Emotional\\Regulation (ER)}}
 & 30\% & 0.831 & 0.583 & 0.626 \\
 & 60\% & 0.644 & 0.503 & 0.587 \\
 & 100\% & 0.572 & 0.481 & 0.573 \\
\bottomrule
\end{tabular}
\par\vspace{2pt}
{\scriptsize\rmfamily
\textit{Note}: Control denotes the no-intervention baseline. RC = Rational Cognition; EP = Empathy Priming; ER = Emotional Regulation. \textbf{Bold}: best per metric. All experiments start from the same checkpoint.}
\end{table}

The experiment reveals several key findings. First, both RC and ER strategies significantly reduce the Negative Emotion Ratio (NER), showing clear dose-response effects. As shown in Table~\ref{tab:priming}, RC reduces NER from 0.844 to 0.571 (a 32.3\% reduction), with overall emotional intensity declining substantially. Beyond 60\% coverage, RC's effectiveness shows a pronounced threshold jump. ER likewise exhibits significant dose-response effects, reducing NER to 0.572 at 100\% coverage (comparable to RC), revealing the complementarity of affective-level regulation and cognitive-level guidance in emotional mitigation.

\textbf{The empathy paradox.} The most theoretically significant finding concerns the Empathy Priming (EP) strategy. Social psychology research shows that empathy, while fostering understanding of others' suffering, can simultaneously induce empathic distress in the observer, potentially intensifying negative emotion diffusion under certain conditions~\citep{batson_empathy_1997,bloom_against_2016}. In our experiment, EP's NER reaches 0.878, \textit{higher} than the control group's 0.844, and rises monotonically as coverage increases from 30\% to 100\%, showing a reverse dose-response effect opposite to RC and ER. Mechanistically, empathy priming modifies psychological cognition beliefs ($B^{psy}$) to heighten agents' sensitivity to others' emotions, causing them to absorb negative affective coloring more intensely during belief updates. This then propagates through the emotional contagion mechanism across the social network in a positive feedback loop, ultimately accelerating cascading diffusion of empathetic negative sentiment. Social-BDI's cognitive architecture makes this complete causal chain, from cognitive belief modification to emotional state change to social contagion amplification, fully traceable and interpretable.

\begin{figure}[!t]
\centering
\includegraphics[width=\columnwidth]{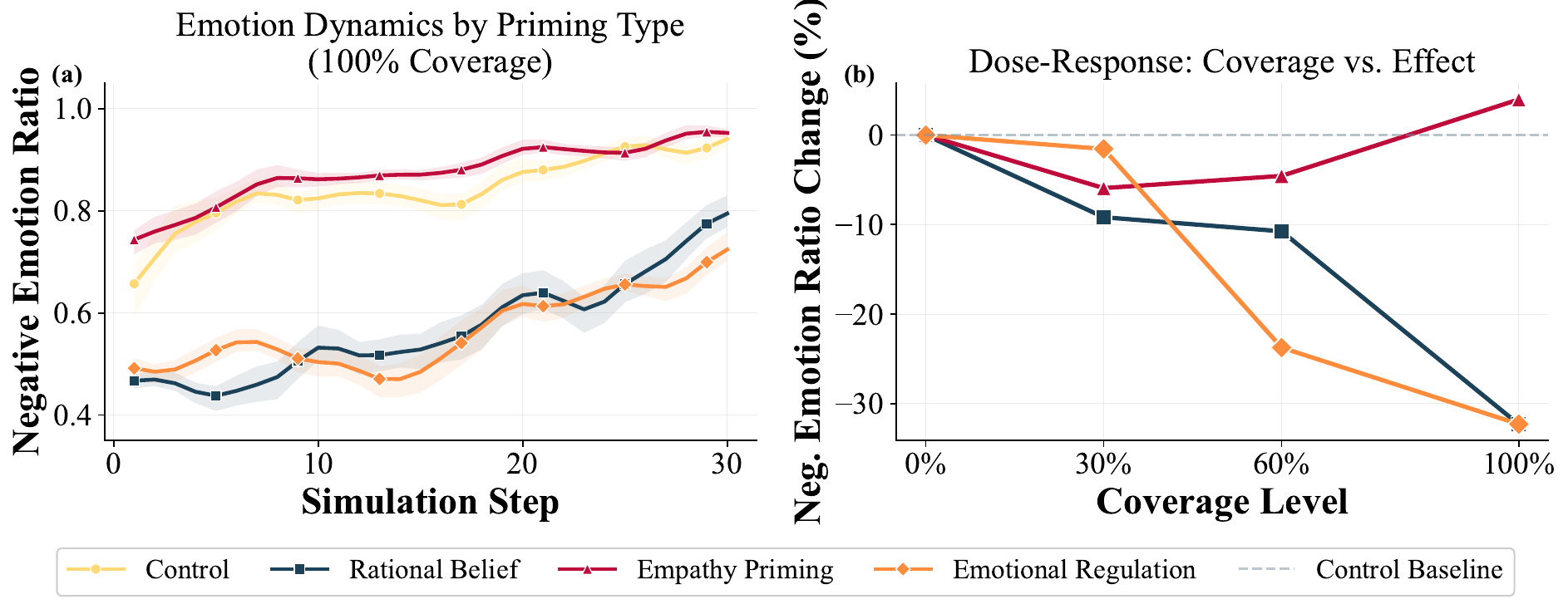}
\caption{Empathy paradox experimental results. (a)~Emotional arousal evolution under two strategies: the empathy priming group (dashed line) surpasses the control group (solid line) from mid-simulation onward; (b)~Behavioral distribution comparison: the empathy group produces a higher proportion of emotion-intensive behaviors (short comments, repost with comment); (c)~Discourse irrationality comparison: the empathy group exhibits a higher proportion of irrational discourse.}
\label{fig:cognitive_priming}
\end{figure}

As shown in Figure~\ref{fig:cognitive_priming}, this finding demonstrates that by directly modifying agents' cognitive belief states, the Social-BDI architecture faithfully propagates belief-level changes through the cognitive pipeline layer by layer, giving rise to complex collective dynamics consistent with social psychology theory. The quantitative reproduction of the empathy paradox shows that even well-intentioned belief adjustments can, through multi-layer cognitive propagation and social network amplification, produce collective effects opposite to expectations. This observation suggests that cognitive-level guidance in public opinion governance requires careful assessment of cascading consequences in multi-agent interaction environments. Compared to heightening empathic sensitivity, directly operating on cognitive judgment (rational cognition priming) or affective experience (emotional regulation) may offer greater controllability.

\subsection{Counterfactual Strategy Evaluation}

Evaluating the effectiveness of corporate crisis communication strategies is inherently irreversible in real settings. Once a response approach is chosen and executed, there is no going back to test alternatives. Counterfactual simulation experiments enable decision-makers to compare multiple options under identical initial conditions before committing, providing direct practical value for crisis response planning.

We design a counterfactual experiment based on the Xibei Prepared Food event, keeping external events fixed while replacing only Xibei's public relations response, comparing five strategies. \textbf{Actual Response} (AR) reproduces Xibei's documented public relations response sequence as recorded in the original data. \textbf{Swift Early Apology} (SEA) issues a sincere apology at the event's onset. \textbf{Proactive Transparency} (PT) voluntarily discloses prepared food usage and food safety reports. \textbf{Consumer Dialogue} (CD) engages in direct interaction and communication with consumers. \textbf{Strategic Silence} (SS) makes no public response. Results are shown in Table~\ref{tab:counter}.

\begin{table}[!t]
\centering
\caption{Counterfactual strategy evaluation results.}
\label{tab:counter}
\scriptsize
\renewcommand{\arraystretch}{1.0}
\begin{tabular}{@{}lccc@{}}
\toprule
\textbf{Strategy} & \makecell{\textbf{Neg.\ Emotion}\\\textbf{Ratio}\,$\downarrow$} & \makecell{\textbf{Anger}\\\textbf{Ratio}\,$\downarrow$} & \makecell{\textbf{Emotion}\\\textbf{Intensity}\,$\downarrow$} \\
\midrule
Actual Response (AR) & 0.792 & 0.685 & 0.685 \\
Swift Early Apology (SEA) & 0.749 & 0.621 & 0.612 \\
Proactive Transparency (PT) & 0.773 & 0.658 & 0.645 \\
\rowcolor{grayrow}
Consumer Dialogue (CD) & \best{0.744} & \best{0.612} & \best{0.598} \\
Strategic Silence (SS) & 0.831 & 0.714 & 0.702 \\
\bottomrule
\end{tabular}
\par\vspace{2pt}
{\scriptsize\rmfamily
\textit{Note}: AR = Actual Response; SEA = Swift Early Apology; PT = Proactive Transparency; CD = Consumer Dialogue; SS = Strategic Silence. \textbf{Bold}: best per metric. All strategies branch from the same pre-crisis checkpoint.}
\end{table}

The experiment reveals several key findings. Proactive strategies significantly outperform passive approaches, with CD yielding the best results (NER = 0.744), followed by SEA (0.749) and PT (0.773). Strategic Silence performs worst, with NER reaching 0.831 and emotional intensity even exceeding the Actual Response, confirming that silence is not the optimal choice in crisis communication. Strategy propagation strictly follows the causal chain of belief revision $\to$ cognitive reasoning change $\to$ behavioral output transformation, validating the interpretability of the Social-BDI architecture in strategy evaluation.

\subsection{Key Findings}

The validation experiments support the following findings. At the individual level, the Social-BDI architecture outperforms all baselines on cognitive-behavior chain consistency, personality stability, and decision robustness, validating the advantage of layered cognitive modeling over end-to-end generation. Notably, CoT achieves the lowest decision robustness, indicating that serialized reasoning within a single call lacks stable state anchoring, whereas Social-BDI provides effective cognitive grounding through independently modularized explicit belief states. At the collective level, the simulation spontaneously gives rise to public opinion lifecycle, behavioral heterogeneity, emotional polarization, scale-free topology, and cascade power-law distributions, all consistent with social science theory, without any explicit group-level programming. At the statistical calibration level, POSIM outperforms baselines on all three datasets, with behavior, content, and topology metrics improving by \textbf{5.0\%}, \textbf{13.0\%}, and \textbf{8.5\%} respectively. Ablation experiments demonstrate functional complementarity among cognitive subsystems, with the desire subsystem contributing most to irrational discourse reproduction and the belief subsystem providing indispensable cognitive grounding for personalized decisions.

\subsection{Limitations}

Despite overall positive results, this study has several limitations. First, the cognitive reasoning quality is tightly coupled with the underlying LLM's capabilities, and the inference cost for large-scale simulation still has room for reduction. Second, follower relationships in the social network remain static during simulation, and dynamic follow/unfollow mechanisms have not yet been incorporated. Third, experimental validation covers only the Weibo platform, and the framework's transferability to different social media platforms and cultural contexts requires further investigation. Finally, LLM-based simulation generally requires substantial computational resources, a challenge shared by this work. Balancing token consumption and simulation performance will be a key focus of future research.

\section{Conclusion}\label{sec:conclusion}

``\textit{All models are wrong, but some are useful.}''~\citep{box_draper_1987_empirical_modelbuilding} The key to public opinion simulation lies not in pursuing point-by-point replication of reality, but in obtaining approximations of opinion mechanisms and evolutionary dynamics that are \textbf{sufficiently credible and actionable for decision-making}. This paper presents POSIM, a social media public opinion simulation framework that embeds large language models within the Social-BDI cognitive architecture to construct agents with explicit cognitive states, and unifies exogenous event shocks and endogenous user interactions via the Hawkes self-exciting point process to achieve high-fidelity public opinion evolution simulation at minute-level temporal resolution.

The three-tier progressive validation spanning mechanism, phenomenon, and statistics demonstrates that the Social-BDI architecture outperforms direct prompting and chain-of-thought methods in agent decision quality. The simulation spontaneously gives rise to public opinion lifecycle, emotional polarization, and scale-free topology, all collective behavioral patterns consistent with social science theory. Behavior, content, and topology metrics improve by 5.0\%, 13.0\%, and 8.5\% respectively over the best baselines on three real-world opinion datasets. Two governance-oriented case studies further reveal counterintuitive phenomena such as the empathy paradox, demonstrating that POSIM not only credibly reproduces public opinion evolution but also serves as a practical computational experimentation platform for public opinion governance.

\section*{Ethical Statement}
This study is conducted purely as data-driven scientific research aimed at advancing computational methods for public opinion simulation. All events are analyzed based entirely on publicly available data, and the authors hold no opinions, judgments, or positions regarding any events, individuals, or organizations involved. All descriptions present only publicly documented facts without expressing or implying any evaluative stance. The simulation framework is intended exclusively for academic research and methodological validation.

\section*{Acknowledgments}

This research was supported by the National Defense Science and Technology Key Excellent Fund Project.

\bibliographystyle{plainnat}
\bibliography{posim_refs}

\appendix

\section{Core Prompt Templates}
\label{app:prompts}

This appendix presents the three core prompt templates used within the Social-BDI cognitive architecture (simplified for presentation). Variables in curly braces (e.g., \texttt{\{agent\_role\}}) are dynamically populated at runtime based on agent type and simulation state. Role-specific elements such as cognitive bias factors, desire type lists, behavioral characteristics, and expression style options are parameterized per agent category (Ordinary User, Opinion Leader, Media Account, Government).

\begin{tcolorbox}[colback=promptbg, colframe=promptborder, boxrule=0.5pt, arc=1mm,
 left=6pt, right=6pt, top=5pt, bottom=5pt, enhanced, breakable,
 title={\footnotesize\textbf{A.1~~Belief Update Prompt}}]
\footnotesize
You are currently \texttt{\{agent\_role\}} on the Weibo platform, participating in a public opinion event discussion.

Current event background: \texttt{\{event\_background\}}

\medskip
\textbf{\#\# Belief State Inference}

Current time: \texttt{\{current\_time\}}

Based on the following information, infer your current belief state:

\smallskip
\textbf{\#\#\# My Identity Information:} \texttt{\{identity\_text\}}

\textbf{\#\#\# My Previous Belief State:} \texttt{\{belief\_text\}}

\textbf{\#\#\# Newly Received Information:} \texttt{\{new\_info\}}

\textbf{\#\#\# My Historical Behavior Memory:} \texttt{\{memories\}}

\textbf{\#\#\# External Events:} \texttt{\{external\_events\}}

\smallskip
Please consider the following cognitive factors:

\texttt{\{cognitive\_factors\}}

{\scriptsize\textit{// e.g., for Ordinary User: confirmation bias, anchoring effect, emotional contagion, herd mentality, emotion-first reasoning, simplistic attribution; for Opinion Leader: brand image, professional judgment, public opinion trend, self-reinforcement.}}

\smallskip
Output your current belief state in JSON format:
{\scriptsize\begin{lstlisting}
{
  "psychological_cognition": ["cognition_trait_1", ...],
  "event_opinions": [
    {"subject": "entity_name", "opinion": "opinion_text"}
  ],
  "emotion_vector": {
    "happy": 0.0, "sad": 0.0, "angry": 0.0,
    "fear": 0.0, "surprise": 0.0, "disgust": 0.0
  }
}
\end{lstlisting}
}
{\scriptsize\textit{Note}: Each emotion dimension ranges from 0 (absent) to 1 (extreme).}
\end{tcolorbox}

\begin{tcolorbox}[colback=promptbg, colframe=promptborder, boxrule=0.5pt, arc=1mm,
 left=6pt, right=6pt, top=5pt, bottom=5pt, enhanced, breakable,
 title={\footnotesize\textbf{A.2~~Desire Generation Prompt}}]
\footnotesize
You are \texttt{\{agent\_role\}} on the Weibo platform. Based on your current beliefs and information you have seen, infer your current social desires.

Current event background: \texttt{\{event\_background\}}

\medskip
\textbf{\#\# Desire Reasoning}

Current time: \texttt{\{current\_time\}}

\smallskip
\textbf{\#\#\# My Current Belief State:} \texttt{\{belief\_text\}}

\textbf{\#\#\# Information I Have Seen:} \texttt{\{exposed\_info\}}

\textbf{\#\#\# External Events:} \texttt{\{external\_events\}}

\smallskip
\textbf{Desire Types} (prioritized for \texttt{\{agent\_role\}}):

\texttt{\{desire\_types\}}

{\scriptsize\textit{// e.g., for Ordinary User: emotional venting, entertainment, self-expression, social approval, social connection, justice advocacy, information seeking; for Opinion Leader: influencing others, self-expression, social approval, justice advocacy, information seeking.}}

\smallskip
Output the desire list in JSON format:
{\scriptsize\begin{lstlisting}
{
  "desires": [
    {"type": "desire_type",
     "description": "specific desire description",
     "intensity": "very_low/low/medium/high/very_high"}
  ]
}
\end{lstlisting}
}
\end{tcolorbox}

\begin{tcolorbox}[colback=promptbg, colframe=promptborder, boxrule=0.5pt, arc=1mm,
 left=6pt, right=6pt, top=5pt, bottom=5pt, enhanced, breakable,
 title={\footnotesize\textbf{A.3~~Intention \& Action Decision Prompt}}]
\footnotesize
You are currently \texttt{\{agent\_role\}} on the Weibo platform. Combine your beliefs, desires, and breaking events to decide actions.

Current event background: \texttt{\{event\_background\}}

\smallskip
\textbf{[Behavioral Characteristics]}

\texttt{\{behavioral\_characteristics\}}

{\scriptsize\textit{// e.g., for Ordinary User: colloquial language, emotional expression, internet slang, exaggerated and biased allowed; for Opinion Leader: original viewpoints preferred, in-depth content, influence-oriented.}}

\medskip
\textbf{\#\# Action--Intention Decision}

Current time: \texttt{\{current\_time\}}

\textbf{\#\#\# My Current Beliefs:} \texttt{\{belief\_text\}}

\textbf{\#\#\# My Current Desires:} \texttt{\{desires\}}

\textbf{\#\#\# Available Posts (indexed):} \texttt{\{exposed\_posts\}}

\texttt{\{hot\_topics\}} \quad \texttt{\{external\_events\}}

\smallskip
\textbf{[Decision Guidelines]}

Determine action quantity and type based on: event heat, emotional intensity, relevance, temporal decay, exposure content, and behavioral tendency described in agent identity.

\smallskip
\textbf{Three-level chain-of-thought reasoning:}

\textbf{L1: Action type \& target:} Choose among \textit{Like}, \textit{Repost}, \textit{Repost w/ comment}, \textit{Short comment}, \textit{Long comment}, \textit{Short post}, \textit{Long post}; specify target post index (0 for original).

\textbf{L2: Expression strategy} (skip for Like/Repost):
\begin{itemize}[nosep, leftmargin=1.5em]
\item Emotion type \& intensity, Stance \& strength, Expression style, Narrative strategy
\item \texttt{\{expression\_styles\}}, \texttt{\{narrative\_strategies\}}
\end{itemize}

{\scriptsize\textit{// e.g., for Ordinary User styles: sarcastic, aggressive, emotional venting, mocking; for Opinion Leader: rational, ironic, empathetic, questioning.}}

\textbf{L3: Content generation} (skip for Like/Repost): Generate Weibo text matching \texttt{\{agent\_role\}} voice, consistent with L1 and L2 decisions.

\smallskip
Output the action list in JSON format:
{\scriptsize\begin{lstlisting}
{
  "actions": [
    {
      "action_type": "short_comment",
      "target_id": 3,
      "expression_strategy": {
        "emotion_type": "anger", "emotion_intensity": "high",
        "stance": "oppose_X", "stance_intensity": "high",
        "style": "sarcastic", "narrative": "labeling"
      },
      "content": {
        "text": "Generated Weibo text...",
        "topics": ["#topic#"], "mentions": ["@user"]
      }
    }
  ]
}
\end{lstlisting}
}
{\scriptsize\textit{Note}: Like/Repost actions omit expression strategy and content fields. Return empty action list if no action is taken.}
\end{tcolorbox}

\section{Evaluation Prompt Templates}
\label{app:eval_prompts}

This appendix presents the three LLM-as-Judge evaluation prompt templates used in the micro-level behavioral mechanism validation (Section~\ref{subsec:micro}). Each template is shown in simplified form; the actual prompts include additional formatting instructions and output schema constraints.

\begin{tcolorbox}[colback=promptbg, colframe=promptborder, boxrule=0.5pt, arc=1mm,
 left=6pt, right=6pt, top=5pt, bottom=5pt, enhanced, breakable,
 title={\footnotesize\textbf{B.1~~Cognitive-Behavior Chain Consistency Evaluation}}]
\footnotesize
You are an expert in social psychology and cognitive science. Please analyze the following social media user's cognitive-behavioral data and evaluate the rationality and cognitive depth of their behavioral decision-making process from a cognitive architecture perspective.

User identity: \texttt{\{identity\}}

Psychological beliefs: \texttt{\{beliefs\}}

\texttt{\{cognitive\_chain\}}

\smallskip
Please evaluate on the following four sub-dimensions, providing an integer rating from 0--5 with brief justification for each:
\begin{enumerate}[nosep, leftmargin=1.5em]
\item \textbf{Cognitive chain completeness}: Does the decision process reflect a complete, verifiable cognitive reasoning chain from perception through belief, desire, and intention to final action?
\item \textbf{Emotional dynamics rationality}: Do cross-round emotional changes conform to cognitive psychology principles (e.g., gradual decay in the absence of stimuli, intensification upon exposure to provocative content)?
\item \textbf{Motivation-behavior alignment}: Can the observed behaviors be traced back to independently modeled cognitive motivations? Is the mapping from desire weights to action types logically consistent?
\item \textbf{Information integration}: Does the cognitive state exhibit an independent update mechanism when facing new information, rather than simply overwriting previous states?
\end{enumerate}
\end{tcolorbox}

\begin{tcolorbox}[colback=promptbg, colframe=promptborder, boxrule=0.5pt, arc=1mm,
 left=6pt, right=6pt, top=5pt, bottom=5pt, enhanced, breakable,
 title={\footnotesize\textbf{B.2~~Personality Stability Evaluation}}]
\footnotesize

This evaluation follows a two-step protocol: first inferring personality traits from behavioral data alone, then comparing the inferred profile against the ground-truth profile used to initialize the agent.

\medskip
\textbf{Step 1: Personality Inference.}

You are a personality psychology expert. Based on the following social media user's complete behavioral and cognitive trajectory across \texttt{\{num\_rounds\}} simulation rounds, infer their personality traits.

\smallskip
\textbf{Input data:}
\begin{itemize}[nosep, leftmargin=1.5em]
\item All posts, comments, and reposts generated by the agent
\item Emotional state trajectory across rounds
\item Interaction patterns (reply targets, repost sources)
\end{itemize}

\texttt{\{user\_data\}}

\smallskip
\textbf{Please infer the following structured profile:}
\begin{enumerate}[nosep, leftmargin=1.5em]
\item \textbf{Identity description}: A first-person narrative summarizing the user's self-perceived social role, values, and communication style.
\item \textbf{Psychological beliefs} (3--5 items): Core cognitive tendencies that drive the user's information processing and opinion formation.
\item \textbf{Behavioral tendencies}: Characteristic patterns in content generation, interaction preferences, and emotional expression.
\item \textbf{Emotional characteristics}: Six-dimensional emotion baseline (anger, sadness, fear, disgust, surprise, joy) estimated from behavioral signals.
\end{enumerate}

\medskip
\textbf{Step 2: Similarity Assessment.}

Please evaluate the similarity (0--1) between the following two personality profiles.

\smallskip
\textbf{[Original personality profile]}

\quad Identity: \texttt{\{original\_identity\}}

\quad Psychological beliefs: \texttt{\{original\_beliefs\}}

\quad Behavioral tendencies: \texttt{\{original\_tendency\}}

\smallskip
\textbf{[Behavior-inferred personality profile]}

\quad Identity: \texttt{\{inferred\_identity\}}

\quad Psychological beliefs: \texttt{\{inferred\_beliefs\}}

\quad Behavioral tendencies: \texttt{\{inferred\_tendency\}}

\smallskip
\textbf{Evaluation dimensions} (each scored 0--1):
\begin{enumerate}[nosep, leftmargin=1.5em]
\item \textbf{Identity consistency}: Does the inferred self-description match the original in terms of social role, values, and communication style?
\item \textbf{Psychological trait consistency}: Do the inferred cognitive tendencies align with the original belief system?
\item \textbf{Behavioral tendency consistency}: Are the inferred behavioral patterns consistent with the original profile?
\end{enumerate}
\end{tcolorbox}

\begin{tcolorbox}[colback=promptbg, colframe=promptborder, boxrule=0.5pt, arc=1mm,
 left=6pt, right=6pt, top=5pt, bottom=5pt, enhanced, breakable,
 title={\footnotesize\textbf{B.3~~Decision Robustness Evaluation}}]
\footnotesize

You are a cognitive psychology expert evaluating the cognitive robustness of a social media user simulation system. The agent made multiple decisions under semantically equivalent but superficially perturbed variants of the same scenario.

\smallskip
\textbf{Perturbation types:}
\begin{itemize}[nosep, leftmargin=1.5em]
\item Reordering of recommended posts in the agent's feed
\item Addition or removal of 1--2 posts from the recommendation list
\item Minor paraphrasing of peripheral context information
\end{itemize}

\medskip
\textbf{Agent information:}

\quad Agent identity: \texttt{\{identity\}}

\quad Cognitive architecture type: \texttt{\{method\_label\}}

\medskip
\textbf{Original decision} (under unperturbed input):

\texttt{\{original\_decision\}}

\medskip
\textbf{Decisions under perturbed conditions} (\texttt{\{num\_variants\}} variants):

\texttt{\{perturbed\_decisions\}}

\smallskip
For each perturbed decision, note whether the core behavioral choice (action type, target, and emotional tone) changed relative to the original.

\medskip
\textbf{Evaluation dimensions} (each scored 0--1):
\begin{enumerate}[nosep, leftmargin=1.5em]
\item \textbf{Core decision stability}: Does the core behavioral decision (action type and primary target) remain stable under minor input perturbations? A score of 1 indicates identical core decisions across all variants.
\item \textbf{Cognitive anchoring strength}: Does the decision exhibit anchoring effects driven by stable underlying beliefs and values, rather than being swayed by superficial input variations?
\item \textbf{Adaptive rationality}: When differences exist between the original and perturbed decisions, are they reasonable cognitive adaptations to meaningful input changes, or do they appear as random noise?
\end{enumerate}
\end{tcolorbox}

\FloatBarrier
\section{Parameter Configuration}
\label{app:params}

POSIM's temporal dynamics and recommendation mechanisms rely on key parameters governing agent activation and content exposure. Table~\ref{tab:hawkes_params} presents Hawkes process parameters controlling temporal evolution, and Table~\ref{tab:rec_params} lists recommendation parameters balancing homophily, popularity, and freshness. These parameters were calibrated through iterative experiments with manual validation, demonstrating robust generalizability across all three events.

\begin{table}[!htb]
\centering
\caption{Hawkes point process parameters (unified across three events).}
\label{tab:hawkes_params}
\small
\renewcommand{\arraystretch}{1.15}
\begin{tabular}{@{}lcc@{}}
\toprule
\textbf{Parameter} & \textbf{Symbol} & \textbf{Value} \\
\midrule
Background rate & $\mu$ & 0.01 \\
Endogenous excitation & $\alpha_{int}$ & 0.005 \\
Endogenous decay & $\beta_{int}$ & 0.16 \\
Exogenous excitation & $\alpha_{ext}$ & 0.08 \\
Exogenous decay & $\beta_{ext}$ & 0.005 \\
Scaling factor & $N_{s}$ & 2000 \\
Circadian amplitude & $s_{circ}$ & 0.3 \\
\bottomrule
\end{tabular}
\end{table}

\begin{table}[!htb]
\centering
\caption{Recommendation system parameters.}
\label{tab:rec_params}
\small
\renewcommand{\arraystretch}{1.15}
\begin{tabular}{@{}lcc@{}}
\toprule
\textbf{Parameter} & \textbf{Symbol} & \textbf{Value} \\
\midrule
Homophily weight & $w_h$ & 0.3 \\
Popularity weight & $w_p$ & 0.3 \\
Freshness weight & $w_r$ & 0.4 \\
Exploration rate & $r_{explore}$ & 0.2 \\
Rec.\ count & $K$ & 10 \\
Comment sampling & $N_c$ & 5 \\
Embedding model & --- & BGE-small-zh \\
\bottomrule
\end{tabular}
\end{table}

\section{Data Collection and Preprocessing}
\label{app:data}

The three public opinion event datasets used in this study are all collected from the Sina Weibo platform. For each event, the collection scope includes original posts containing event-related hashtags or keywords along with their complete repost chains and comment trees, as well as basic information of all participating users (username, gender, region, verification type, follower count, personal description, etc.). The raw data volumes are: LE comprises 426,882 users and 1,254,512 behavioral records; WL comprises 416,235 users and 1,416,034 records; XF comprises 725,251 users and 1,851,124 records. In total, the three datasets encompass 1,568,368 users and 4,521,670 behavioral records.

The raw data undergoes multi-stage cleaning: deduplication based on post IDs; filtering of low-activity users with insufficient posts during the simulation period, ensuring each agent has adequate behavioral samples for belief initialization; removal of advertisements and spam via keyword blacklists; filtering of content irrelevant to the event topic; and quality filtering based on content length. After this cleaning pipeline, the retained data volumes are: 1,530 users and 34,218 posts for LE; 1,843 users and 51,647 posts for WL; 1,987 users and 14,892 posts for XF (see Table~\ref{tab:dataset}).

Agent belief initialization proceeds as follows: role identity beliefs from users' basic information and historical content; psychological cognition beliefs matched from a predefined trait pool; event opinion beliefs extracted through LLM-driven structured interviews; and initial emotion vectors from sentiment analysis of recent posts.

\section{Computational Resource Consumption}
\label{app:efficiency}

Table~\ref{tab:efficiency} summarizes the computational resource consumption for the three event simulations. All LLM inference is performed via the SiliconFlow API\footnote{\url{https://siliconflow.cn/}}.

\noindent
\begin{minipage}[t]{\columnwidth}
\centering
\captionof{table}{Computational resource consumption for three event simulations.}
\label{tab:efficiency}
\small
\renewcommand{\arraystretch}{1.15}
\begin{tabular*}{\columnwidth}{@{\extracolsep{\fill}}lrrr@{}}
\toprule
\textbf{Metric} & \textbf{LE} & \textbf{WL} & \textbf{XF} \\
\midrule
No.\ of agents & 1,530 & 1,843 & 1,987 \\
Simulation steps & 276 & 1,140 & 426 \\
Simulated duration & $\sim$46\,h & $\sim$190\,h & $\sim$71\,h \\
Total actions & 29,146 & 46,140 & 12,261 \\
LLM API calls & 26,295 & 39,945 & 11,670 \\
Total tokens & 28.98\,M & 46.95\,M & 13.37\,M \\
Wall-clock time & $\sim$3.5\,h & $\sim$8.3\,h & $\sim$1.9\,h \\
\bottomrule
\end{tabular*}
\end{minipage}

\end{document}